\documentclass[11pt,fleqn]{article}
\usepackage{amsmath}
\usepackage{amssymb} 
\usepackage{mathtools}
\usepackage[margin=1in,bottom=1.25in,top=1.25in]{geometry}
\usepackage{newtxtext}
\usepackage{newtxmath}
\usepackage{dsfont} 
\usepackage{caption} 
\usepackage{subcaption}
\usepackage{mathrsfs}	
\usepackage{cite} 
\usepackage{enumitem}
\usepackage{array}
\usepackage{booktabs}
\usepackage[math]{cellspace}
\usepackage{wrapfig}
\usepackage{cancel}
\usepackage{nicefrac}
\usepackage{pifont} 
\usepackage[linkcolor=blue,urlcolor=blue,citecolor=magenta,colorlinks=true]{hyperref}
\usepackage{cleveref}
\usepackage[bottom]{footmisc}
\usepackage{soul}
\usepackage[skins,breakable,theorems]{tcolorbox}
\usepackage[nolist]{acronym} 
\usepackage{braket}
\usepackage{mleftright}\mleftright
\usepackage{accents}
\usepackage{xfrac}
\usepackage{xspace}
\usepackage{eqparbox}
\usepackage{tikz}
\usetikzlibrary{positioning} 
\makeatletter
\NewDocumentCommand{\eqmathbox}{o O{c} m}{%
  \IfValueTF{#1}
    {\def\eqmathbox@##1##2{\eqmakebox[#1][#2]{$##1##2$}}}
    {\def\eqmathbox@##1##2{\eqmakebox{$##1##2$}}}
  \mathpalette\eqmathbox@{#3}
}
\makeatother

\makeatletter
\g@addto@macro\bfseries{\boldmath}
\newcommand*{\defeq}{\mathchoice{\mathrel{\rlap{%
\raisebox{0.24ex}{$\m@th\cdot$}}%
\raisebox{-0.24ex}{$\m@th\cdot$}}%
=}{\mathrel{\rlap{%
\raisebox{0.24ex}{$\m@th\cdot$}}%
\raisebox{-0.24ex}{$\m@th\cdot$}}%
=}{\mathrel{\rlap{%
\raisebox{0.08ex}{\small$\m@th\cdot$}}%
\raisebox{-0.28ex}{\small$\m@th\cdot$}}%
=}{\mathrel{\rlap{%
\raisebox{0.08ex}{\tiny$\m@th\cdot$}}%
\raisebox{-0.28ex}{\tiny$\m@th\cdot$}}%
=}}
\newcommand*{\eqdef}{\mathchoice{=\mathrel{\rlap{%
\raisebox{0.24ex}{$\m@th\cdot$}}%
\raisebox{-0.24ex}{$\m@th\cdot$}}}{%
=\mathrel{\rlap{%
\raisebox{0.24ex}{$\m@th\cdot$}}%
\raisebox{-0.24ex}{$\m@th\cdot$}}}{%
=\mathrel{\rlap{%
\raisebox{0.08ex}{\small$\m@th\cdot$}}%
\raisebox{-0.28ex}{\small$\m@th\cdot$}}}{%
=\mathrel{\rlap{%
\raisebox{0.08ex}{\tiny$\m@th\cdot$}}%
\raisebox{-0.28ex}{\tiny$\m@th\cdot$}}}%
}
\newcommand*{\transpose}{
    {\mathpalette\@transpose{}}%
    }
\newcommand*{\@transpose}[2]{%
    \raisebox{\depth}{$\m@th#1\intercal$}%
}
\makeatother
\allowdisplaybreaks
\newcommand{\ii}{\mathop{}\!\mathrm{i}\!\mathop{}}
\newcommand{\ee}{\mathrm{e}}
\newcommand\simnsam{\mathrel{\ooalign{$\simeq$\cr
  \hidewidth\raise-.333ex\hbox{\rotatebox{45}{$\shortmid$}}\hidewidth\cr}}}
\DeclareMathOperator{\re}{Re}
\DeclareMathOperator{\im}{Im}

\DeclareMathOperator{\diag}{diag}

\newcommand{\Group}[3][]{\text{#2}(#3)}

\newcommand{\ChargeC}{\ensuremath{\mathcal{C}}}

\newcommand{\ParityP}{\ensuremath{\mathcal{P}}}
\newcommand{\CP}{\ensuremath{\ChargeC\ParityP}\xspace}

\newcommand{\rep}[1]{\ensuremath{{\boldsymbol{#1}}}}

\newcommand{\GeneratorS}{\mathsf{S}}
\newcommand{\GeneratorT}{\mathsf{T}}
\newcommand{\Boxed}[2][]{\tcboxmath[arc=0mm,boxrule=0pt,left=0pt,right=0pt,top=0pt,bottom=0pt,colback=blue!15!white,oversize,enlarge top by=-2.6pt,enlarge bottom by=-2pt,#1]{#2}}
\newcommand{\TpO}{\ensuremath{\rep{2}\oplus\rep{1}}}
\newcommand{\vev}[1]{\ensuremath{\langle{#1}\rangle}}
\DeclareMathSymbol{\mathordcolon}{\mathord}{operators}{"3A}
\newcommand{\relative}{\mathrel{\mathordcolon}}

\date{}

\begin{document}
\title{
\begin{flushright}
\hfill\mbox{\small\texttt{UCI-TR-2024-22}}\\[5mm]
\begin{minipage}{0.2\linewidth}
\normalsize
\end{minipage}
\end{flushright}
{\Large\bfseries
Modular Flavor Symmetries and Fermion Mass Hierarchies\\[2mm]
}
\author{
Mu-Chun Chen\footnote{E-mail: \texttt{muchunc@uci.edu}},\quad
Xueqi Li\footnote{E-mail: \texttt{xueqi.li@uci.edu}},\quad
Xiang-Gan Liu\footnote{E-mail: \texttt{xianggal@uci.edu}}\quad and\quad 
Michael Ratz\footnote{E-mail: \texttt{mratz@uci.edu}}%
\\*[20pt]
\centerline{
\begin{minipage}{\linewidth}
\begin{center}
{\itshape\small Department of Physics and Astronomy, University of California, Irvine, CA 92697-4575 USA}
\end{center}
\end{minipage}}
\\[10mm]}}

\maketitle
\thispagestyle{empty}

\begin{abstract}
We investigate fermion mass hierarchies in models with modular flavor symmetries. 
Several key conclusions arise from the observation that the determinants of mass matrices transform as 1-dimensional vector-valued modular forms. 
We demonstrate that, under some fairly general assumptions, achieving hierarchical fermion masses requires the vacuum expectation value of the modulus $\tau$ to be located near one of the critical points, $\ii$, $\ii \infty$, or $\omega$. 
We also revisit the universal near-critical behavior around these points and classify the resulting mass hierarchies for the critical points $\ii$ and $\omega$. 
We compare the traditional Froggatt--Nielsen mechanism with its modular variant. 
The knowledge and boundedness of Fourier and Taylor coefficients are crucial to the predictive power of modular flavor symmetries.
\end{abstract}
\clearpage

\section{Introduction}
\label{sec:Introduction}

The flavor puzzle remains as one of the most intriguing unsolved problems in the \ac{SM}. 
While there is still a lack of a universally accepted solution, one avenue that has garnered significant attention is the flavor symmetry approach. 
Under the guiding principle of flavor symmetry, Yukawa couplings are promoted to a set of functions dependent on scalar fields referred to as ``flavons''.  
The \acp{VEV} of these flavons determine the values of the Yukawa couplings, and the flavor puzzle thus becomes a vacuum selection problem. 
In the traditional flavor symmetry framework, achieving realistic predictions for the flavor parameters in the \ac{SM} requires intricate \ac{VEV} alignments (see e.g.\ \cite{Holthausen:2011vd}) involving multiple flavon fields. 
On the other hand, the recently emerging framework of modular flavor symmetries~\cite{Feruglio:2017spp} largely avoids the vacuum alignment problem by assuming that the Yukawa couplings are modular forms. 
Modular forms depend on only one flavon, the so-called modulus $\tau$. 
This greatly simplifies the vacuum alignment problem to the vacuum selection of a single modulus field. 
Furthermore, the modular symmetry constrains Yukawa couplings to a finite set of modular forms, which also dictate the structure of all higher-order operators involving the modulus. 
Therefore, the scheme of modular flavor symmetries has the potential to be highly predictive.
The construction of the mass models hinges on the choice of a finite modular group, $\Gamma_N$~\cite{Feruglio:2017spp}, or the double covering of a finite modular group, $\Gamma_N'$~\cite{Liu:2019khw},  along with the assignments of modular weights and representations for matter fields. 
This results in a wide range of possibilities, as shown in various bottom-up model~\cite{Ding:2019zxk,Novichkov:2018nkm,Ding:2019xna,Qu:2021jdy,Liu:2020akv,Liu:2020msy,Novichkov:2020eep,Yao:2020zml,Wang:2020lxk,Ding:2020msi,Li:2021buv,Meloni:2023aru,Ding:2023ydy}.\footnote{For further references, see the recent reviews~\cite{Feruglio:2019ybq,Almumin:2022rml,Kobayashi:2023zzc,Ding:2023htn,Ding:2024ozt}.}

It is instructive to examine the locations of $\tau$ \acp{VEV} in bottom-up models. 
It turns out that they tend to be statistically close to certain special points in moduli space.
A recent study~\cite{Feruglio:2023bav,Feruglio:2023mii,Ding:2024xhz} suggests that this proximity implies a universal near-critical behavior in modular flavor models. 
In particular, a large class of models exhibits the same power law behavior when the modulus approaches a fixed point. 
In~\cite{Feruglio:2023bav,Feruglio:2023mii}, the three generations of left-handed leptons $L$ are assumed to form an irreducible triplet of $\Group{SL}{2,\mathds{Z}}$. 
With this assignment, a promising neutrino mass pattern is obtained when the modulus is close to the critical point $\tau = \ii$.  
This result agrees with the observation that most models with an irreducible triplet representation of left-handed lepton $L$ gives a best-fit value around $\tau = \ii$. 
It raises the intriguing possibility that all flavor parameters within the \ac{SM} could potentially originate from near-critical vacua.
The question of why $\tau$ settles close to a critical point is, of course, nothing but a version of the  well-known moduli stabilization problem. 
The modulus $\tau$ can be stabilized successfully close to the critical points~\cite{Dent:2001ut,Kobayashi:2019xvz,Novichkov:2022wvg}, and it has been pointed out that the departure of $\tau$ from the critical value may be related to \ac{SUSY} breaking~\cite{Knapp-Perez:2023nty}.
In top-down models, winding modes, which have the interpretation of gauge fields, become massless at certain critical points~\cite{Li:2025bsr}.

The purpose of this paper is to provide a general discussion of the relation between mass hierarchies and the proximity of $\tau$ to critical points. 
A central observation is that modular symmetry constrains the determinant of the mass matrix to be a specific modular form, which necessarily vanishes at certain locations in the fundamental domain. 
Unless the modular forms describing the fermion masses have large weight, these zeros can only occur at the three critical points (i.e.\ fixed points): $\ii$, $\omega\defeq \ee^{2\pi\ii /3}$ and $\ii\infty$. 
This means that hierarchical fermion masses can naturally emerge if $\tau$ takes a value that is parametrically close to one of these points. 
As we shall demonstrate, the analysis of the determinants leads to powerful constraints on the selection of the finite modular group for flavor models, thus narrowing the range of viable models that can reproduce realistic fermion masses. 

We analyze the the structure of mass matrices in the vicinity of the critical points. 
We show that, under the residual symmetry of the modular group, the mass matrix can be obtained by linearly realizing the residual group. 
Since there are only a limited number of representation decompositions of the matter fields under the residual group, it is possible to fully classify the patterns of mass matrices near the critical points. 

In our work, we also extend the analysis in references~\cite{Feruglio:2023bav,Feruglio:2023mii} by relaxing the assumption that the left-handed lepton $L$ is an irreducible triplet of the modular flavor symmetry. 
Specifically, we consider the $\TpO$ scheme. 
This assignment appears to be closer to what one obtains in explicit string models~\cite{Lebedev:2006kn}, as stressed more recently in \cite{Baur:2019iai,Baur:2020jwc}. We find that, in the vicinity of $\omega$, a pattern consistent with experimental data can emerge.

The rest of this paper is organized as follows.  
In \Cref{sec:framework}, we review the formalism of modular flavor symmetry and establish key properties of 1-dimensional \acp{VVMF} in $\Group{SL}{2,\mathds{Z}}$. 
In \Cref{sec:Zeros_and_mass_hierarchies}, we demonstrate that determinants of mass matrices are 1-dimensional \acp{VVMF}, and show how their zeros provide necessary conditions for generating hierarchical fermion masses.  
In \Cref{sec:critical}, we revisit the universal near-critical behavior, showing how the stabilizer subgroups determine the possible mass matrix patterns. We also extend previous analyses beyond irreducible triplet representations to the $\TpO$ case.
In \Cref{sec:Discussion}, we discuss the implications of our findings, including the limitations of our near-critical analysis and its comparison with the traditional \ac{FN} scheme. \Cref{sec:conclusions} contains our summary. 
In \Cref{app:1and2irreps} we provide a catalogue of the 1- and 2-dimensional representations of $\Group{SL}{2,\mathds{Z}}$, 
\Cref{app:classification} contains a classification of lepton mass matrix patterns for $\TpO$ assignments near critical points, 
and \Cref{app:Estimate} derives estimated upper bounds on the Taylor and Fourier expansion coefficients of modular forms.

\section{Modular symmetry framework}
\label{sec:framework}

In this section, we briefly review the modular flavor symmetry framework and summarize some basic though important properties of 1-dimensional \acp{VVMF}.

\subsection{Modular symmetries and modular forms}

The key ingredient of this framework is a modular symmetry.
In this study, we will mainly focus on $\Group{SL}{2,\mathds{Z}}$, which is generated by
\begin{equation}
    \GeneratorS \defeq 
        \begin{pmatrix}
        0  & 1\\
        -1 & 0
        \end{pmatrix}\quad \text{and}\quad
    \GeneratorT\defeq
        \begin{pmatrix}
        1 & 1\\
        0 & 1
        \end{pmatrix}\;.
\end{equation}
Modular invariant models of flavor host the modulus $\tau$ and additional chiral supermultiplets $\varphi^i$.
The scalar component of $\tau$ takes values in the upper complex half-plane $\mathcal{H}=\left\{\tau\in\mathds{C}\mid\im\tau>0 \right\}$.
Given an element 
\begin{equation}
  \gamma = \begin{pmatrix}
a & b\\
c & d
\end{pmatrix}~\in \Group{SL}{2,\mathds{Z}}\;,
\end{equation}
the modulus $\tau$ and matter fields $\varphi$ transform as 
\begin{subequations}\label{eq:modulartransf}
\begin{align}
   \tau&\xmapsto{~\gamma~}{\frac{a\,\tau + b}{c\,\tau + d}}\;,\\
   \varphi^i&\xmapsto{~\gamma~}{(c\,\tau + d)^{-k_{\varphi^i}}}\,\rho_{\varphi^i}(\gamma)\,\varphi^i\;,
\end{align}
\end{subequations} 
where $\rho_{\varphi^{i}}$ is the representation of the modular group $\Group{SL}{2,\mathds{Z}}$, and $k_{\varphi^i}$ is the modular weight of the field $\varphi^i$. 

As the theory is invariant under certain modular transformations, $\tau$ can be restricted to some part of $\mathcal{H}$, such as the fundamental domain $\mathcal{F}$ of $\Group{SL}{2,\mathds{Z}}$, which can be chosen as the region where $\lvert\tau\rvert \geq 1$ and $\lvert\re(\tau)\rvert \leq 1/2$.
In the fundamental domain, no two points are related by a modular transformation, and any point in $\mathcal{H}$ can be mapped to a point in the fundamental domain via such a transformation.

In this work, most of our attention is on the case in which $\rho_{\varphi^i}$ is a representation with a finite image, meaning the kernel of the representation, $\mathrm{N}$, is a normal subgroup of the modular group $\Group{SL}{2,\mathds{Z}}$. 
In this case, $\rho_{\varphi^i}$ is also a representation of the finite modular group $\Group{SL}{2,\mathds{Z}}/\mathrm{N}$~\cite{Feruglio:2017spp,Liu:2021gwa}.

In the framework with $\mathcal{N}=1$ local \ac{SUSY}, the theory depends on the K\"ahler function
\begin{equation}\label{eq:G-function}
  \mathscr{G}(\tau,\varphi,\bar\tau,\bar\varphi) 
  = 
  K(\tau,\varphi,\bar\tau,\bar\varphi) + \ln\mathscr{W}(\tau,\varphi) + \ln\overline{\mathscr{W}}(\bar\tau,\bar\varphi)\;,
\end{equation}
The K\"ahler potential and superpotential are given by
\begin{subequations}
\begin{align}
   K &= -k_{\mathscr{W}}\ln\bigl(-\ii(\tau - \bar\tau)\bigr) 
   +\sum_i \bigl(-\ii(\tau-\bar\tau)\bigr)^{-k_i}\,\lvert\varphi^i\rvert^2 + \dots\;,\\
   \mathscr{W} &= Y_{ijk}(\tau)\,\varphi^i\,\varphi^j\,\varphi^k + \dots\;. \label{eq:generalSuperpotential}
\end{align}
\end{subequations}
The K\"ahler potential with the terms explicitly shown above is often referred to as the minimal form, $K_{\mathrm{min}} = -k_{\mathscr{W}}\ln\bigl(-\ii(\tau - \bar\tau)\bigr) + \sum_i \bigl(-\ii(\tau-\bar\tau)\bigr)^{-k_i}\lvert\varphi^i\rvert^2$, where the gauge fields and quantum numbers are suppressed. 

In what follows, we will stick to the minimal K\"ahler potential.  
Our results can be generalized to the general, non-minimal case~\cite{Chen:2019ewa}.
We further use the standard supergravity mass units in which the Planck scale is set to unity, $M_\mathrm{P}=1$, and the modular weight of the superpotential, $k_{\mathscr{W}}$, is assumed to be a positive integer. 

In local \ac{SUSY}, $K$ and $\mathscr{W}$ are not individually modular invariant. 
Instead, under the modular transformation~\eqref{eq:modulartransf}, the K\"ahler potential undergoes the K\"ahler transformation
\begin{equation}
 K \xmapsto{~\gamma~} K + k_{\mathscr{W}}\,\bigl(\ln(c\,\tau + d) + \ln(c\, \bar \tau + d)\bigr)\;.
\end{equation}
Modular invariance of K\"ahler function $\mathscr{G}$ in \Cref{eq:G-function} requires the superpotential $\mathscr{W}$ to transform nontrivially,
\begin{equation}
  \mathscr{W}\xmapsto{~\gamma~} (c\,\tau + d)^{-k_{\mathscr{W}}}\, 
  \rho_{\mathscr{W}}(\gamma)\,\mathscr{W}\;.
\end{equation}
Here, $\rho_{\mathscr{W}}$ is a 1-dimensional representation matrix of the modular group, i.e.\ a phase depending on $\gamma$. 
In models with global \ac{SUSY}, $k_{\mathscr{W}}$ is assumed to be zero, i.e.\ the superpotential is modular invariant.

Both in the conventions of bottom-up model building and supergravity, the Yukawa couplings $Y_{ijk}(\tau)$ in \Cref{eq:generalSuperpotential} are \acp{VVMF},
\begin{equation}
  Y_{ijk}(\tau)\xmapsto{~\gamma~} Y_{ijk}(\gamma\tau)=(c\,\tau + d)^{k}\,\rho_Y(\gamma)\,Y_{ijk}(\tau)\;, \; \textnormal{where } k = k_i + k_j + k_k - k_{\mathscr{W}}\;.
\end{equation}
The representation matrices $\rho_Y$, $\rho_{\varphi^i}$, $\rho_{\varphi^j}$ and $\rho_{\varphi^k}$ are required to contract to a 1-dimensional representation matrix $\rho_\mathscr{W}$, i.e.\ a phase.

\subsection{1-dimensional vector-valued modular forms of \texorpdfstring{$\Group{SL}{2,\mathds{Z}}$}{SL(2,Z)}\label{sec:1-dimensional_VVMF}}

In this section, we study 1-dimensional \acp{VVMF} in more detail. 
By definition, a 1-dimensional \ac{VVMF} transforms in a 1-dimensional representation, which may or may not be the trivial singlet.
The modular group $\Group{SL}{2,\mathds{Z}}$ contains twelve 1-dimensional representations, which we denote as $\rep{1}_p$, where $p \in \{0,1,\dots,11\}$. 
They can be obtained from the relations $\GeneratorS^4=(\GeneratorS\,\GeneratorT)^3=\mathds{1}$ satisfied by the generators $\GeneratorS$ and $\GeneratorT$,
\begin{equation}\label{eq:SL2ZsingletsMain}
	\rep{1}_p\colon\quad	\rho_{\rep{1}_p}(\GeneratorS)=\ii^p\;,\qquad \rho_{\rep{1}_p}(\GeneratorT)=\ee^{\pi\ii p/6}\;,\qquad \rho_{\rep{1}_p}(\GeneratorS\,\GeneratorT) = \omega^p\;,
\end{equation}
where the trivial representation $\rep{1}$ is $\rep{1}_0$, and $\omega\defeq\ee^{2\pi\ii/3}$ is the third root of unity.

All 1-dimensional \acp{VVMF} of weight $k$ are known to form a finite-dimensional linear space $\mathcal{M}^{(\textnormal{1D})}_k\bigl(\Group{SL}{2,\mathds{Z}}\bigr)$, generated by $\eta^2$, $E_4$  and $E_6$~\cite{Liu:2021gwa},
\begin{equation}
  \mathcal{M}^{(\textnormal{1D})}_k\bigl(\Group{SL}{2,\mathds{Z}}\bigr) 
  = \sum_{\substack{a+b+c=k\\a,b,c\geq 0}} \alpha_{abc}\,\eta^{2c}\,E_4^a\, E_6^b \;.\label{eq:1VVMF}
\end{equation}
Here, $a,b,c\in\mathds{N}_0$, $\alpha_{abc}\in \mathds{C}$ are constants, $E_4$ and $E_6$ are the Eisenstein series of weight 4 and 6, respectively, and $\eta$ is the Dedekind eta-function.
These functions can be defined as
\begin{subequations}\label{eq:E4E6EtaDef}
\begin{align}
  E_4(\tau) &= 1 + 240 \sum_{n=1}^\infty \sigma_3(n)\,q^n\;,\label{eq:E4E6EtaDef_E_4}\\ 
  E_6(\tau) &= 1 - 504 \sum_{n=1}^\infty \sigma_5(n)\,q^n\;,\label{eq:E4E6EtaDef_E_6}\\
  \eta(\tau) &= q^{1/24}\,\prod_{n=1}^\infty (1-q^n)\;, \label{eq:E4E6EtaDef_eta}
\end{align}
\end{subequations}
where $q \equiv \ee^{2\pi\ii \tau}$ and $\sigma_{p}(n)$ is the divisor sum function.

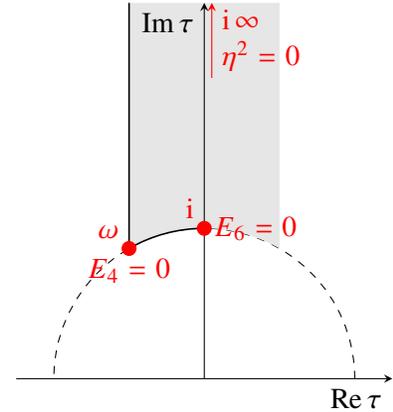
\begin{wrapfigure}[12]{r}{6cm}
\centering\vspace*{-2em}\begin{tikzpicture}[>=stealth,dot/.style={circle,fill,inner sep=2pt},radius=2]
  \draw[dashed] (0:2) arc[start angle=0,end angle=180];
  \path [fill=gray!20] (-1,5) -- (120:2) arc[start angle=120,end angle=60] -- (1,5); 
  \draw[semithick] (-1,5) -- (120:2) arc[start angle=120,end angle=90];
  \draw[->] (-2.5,0) -- (2.5,0) node[below left]{$\re\tau$};
  \draw[->] (0,0) -- (0,5) node[below left]{$\im\tau$};
  \path[red] (120:2) node[dot] {} node[above left]{$\omega$} node[below]{$E_4=0$}
  (90:2) node[dot] {} node[above left]{$\ii$} node[right]{$E_6=0$};
  \draw[red,->] (0.1,4) -- node[right,align=left]{$\ii\infty$\\ $\eta^2=0$} (0.1,5);
\end{tikzpicture}
\caption{Fundamental domain of $\Group{SL}{2,\mathds{Z}}$ and critical points.}
\label{fig:Critical_points}
\end{wrapfigure}

All three functions have unique zeros in the fundamental domain (cf.\ \Cref{fig:Critical_points}),
\begin{subequations}
  \begin{align}
    \eta^2(\tau)=0 &\Longleftrightarrow \tau=\ii\infty\;,\\
    E_4(\tau)=0 &\Longleftrightarrow \tau=\omega\;,\\
    E_6(\tau)=0 &\Longleftrightarrow \tau=\ii\;.
  \end{align}
\end{subequations}
We will refer to these points, $\ii\infty$, $\omega$, and $\ii$, as the critical or fixed points in this paper, with the reasoning explained in more detail in \Cref{sec:critical}.

For generic representations, it can be shown that the lowest weight modular form in representation $\rep{1}_p$ is of the form $\eta^{2p}$ \cite{Liu:2021gwa}. 
Crucially, any higher weight modular form in representation $\rep{1}_p$ can be obtained by taking the product of $\eta^{2p}$ and Eisenstein series $E_4$ and $E_6$, as detailed in \Cref{eq:1VVMF}.

We are interested in the zeros of the 1-dimensional \acp{VVMF}.
As is well known, the zeros of scalar modular forms $f(\tau)$ are governed by the Valence Formula (cf.\ \cite[Proposition 2]{Bruinier:2008xxx})
\begin{equation}
  \frac{1}{12}\,n_{\ii\infty} + \frac{1}{2}\,n_{\ii} + \frac{1}{3}\,n_{\omega} + \sum_{\substack{\tau_0 \neq \\ \ii, \omega, \ii\infty}} n_{\tau_0} = \frac{k}{12}\;.
  \label{eq:ValenceFormula}
\end{equation}
Here, $n_{\ii\infty}$, $n_{\ii}$ and $n_{\omega}$ denote the orders of the zeros at the critical points $\tau=\ii\infty$, $\tau=\ii$ and $\tau=\omega$, respectively, and
the sum extends over the number of zeros away from these points. 
The order of a zero is defined by the asymptotic behavior\footnote{Note that $\tau\mapsto \frac{\tau - \tau_0}{\tau - \bar{\tau}_0}$ is the so-called generalized Cayley transformation, which maps the entire upper half-plane $\mathcal{H}$ to the Poincar\'e unit disk with $\tau_0$ as its origin.
This is the reason why Taylor expansions of modular forms are expansions in $u$ rather than $\tau-\tau_0$ \cite[Section~5]{cohen2017modular}.
Therefore, for  $\tau_0\ne\ii\infty$, $u$ in \eqref{eq:asymptotic} fulfills $0 \leq u \leq 1$ and $u = 0$ when $\tau = \tau_0$.
In all cases, the absolute value $\lvert u\rvert$ serves as a measure of the deviation of $\tau$ from the critical point $\tau_0$.}
\begin{equation}\label{eq:asymptotic}
  f(\tau) \approx c\,u^{n_{\tau_0}}\;,\quad \text{where }  
  u \defeq 
  \begin{dcases}
    \frac{\tau - \tau_0}{\tau - \bar{\tau}_0}&\textnormal{if }\tau_0\neq \ii\infty\;,\\
     \ee^{2\pi\ii\tau/12}&\textnormal{if }\tau_0= \ii\infty\;.
   \end{dcases}
\end{equation}   
Here, $c$ is a constant, and $\im\tau_0>0$.
Note that $n_{\ii\infty}$ must be a nonnegative multiple of 12 for the scalar modular forms of $\Group{SL}{2,\mathds{Z}}$. 
However, as we shall see, this is not true for 1-dimensional \acp{VVMF}.
This is the reason why our conventions for $n_{\ii\infty}$ differ from the ones used in \cite{Bruinier:2008xxx} by a factor 12.

It is possible to generalize the Valence Formula \eqref{eq:ValenceFormula} to the case of nontrivial 1-dimensional \acp{VVMF} of $\Group{SL}{2,\mathds{Z}}$. 
We find that the form remains unchanged.
However, unlike for scalar modular forms, for 1-dimensional \acp{VVMF} $n_{\ii\infty}$ can now be any nonnegative integer, and may not be divisible by 12.

For example, at weight $k=4$, the Valence Formula~\eqref{eq:ValenceFormula} has two nonnegative integer solutions, $(n_{\ii\infty}, n_{\ii},n_{\omega},\sum n_{\tau_0})=(0,0,1,0)$ and $(4,0,0,0)$. 
This is an immediate consequence of the fact that there are only two independent 1-dimensional \acp{VVMF}, $E_4$ and $\eta^8$, at weight 4.
They have only a simple zero at $\omega$ and a zero of order 4 at $\ii\infty$, respectively.

\section{Zeros and mass hierarchies}
\label{sec:Zeros_and_mass_hierarchies}

In this section, we discuss mass hierarchies in the framework of modular flavor symmetries.
Generally, there are various ways of suppressing couplings, such as the \ac{FN} mechanism. 
In this section we focus on scenarios in which modular flavor symmetries are responsible for these hierarchies.

\subsection{Mass matrices and mass hierarchies}
\label{sec:Mass_matrices_and_mass_hierarchies}

Consider a modular invariant bilinear with three generations of matter fields ($i,j\in\{1,2,3\}$),
\begin{equation}\label{eq:superpotential_mass}
 \mathscr{W}=\varphi_i^\ChargeC\, M(\tau)_{ij}\, \varphi_j\;,
\end{equation}
where the matter fields $\varphi^\ChargeC$ and $\varphi$ transform as (cf.\ \Cref{eq:modulartransf})
\begin{subequations}\label{eq:modular_transformation_phi_i_and_phic_i}
\begin{align}
\varphi^\ChargeC_i &\xmapsto{~\gamma~} (c\,\tau+d)^{-k_{\varphi^\ChargeC_i}}\, \rho^\ChargeC_{ij}(\gamma)\,\varphi_j^\ChargeC\;, \\
\varphi_i &\xmapsto{~\gamma~} (c\,\tau+d)^{-k_{\varphi_i}}\,\rho_{ij}(\gamma)\,\varphi_j \;.
\end{align}
\end{subequations}
Notice that this does not require the collection of $\varphi_j$, or $\varphi_i^\ChargeC$, to furnish a single irreducible representation under the modular group.
Rather, \eqref{eq:modular_transformation_phi_i_and_phic_i} also applies to the case of reducible representations.
From \eqref{eq:modular_transformation_phi_i_and_phic_i} one can see that elements of the mass matrix transform as
\begin{equation}
M(\tau)_{ij}\xmapsto{~\gamma~} M(\gamma\tau)_{ij}=(c\,\tau+d)^{k_{\varphi^\ChargeC_i}+k_{\varphi_i}}\,\rho^{\ChargeC\,*}_{ik}\,M(\tau)_{k\ell}\,\rho^{\dagger}_{\ell j}\;.
\end{equation}

The superpotential term \eqref{eq:superpotential_mass} can give rise to Dirac or Majorana masses, depending on whether $\varphi_j$ and $\varphi_i^\ChargeC$ are independent fields. 
It is well known that Dirac and Majorana mass matrices can be diagonalized by unitary transformations,
\begin{subequations}
\begin{align}
 M_\mathrm{Dirac}&=U_\mathrm{L}^\dagger\,\diag(m_1,\dots,m_n)\,U_\mathrm{R}\;,\\
 M_\mathrm{Majorana}&=U^{\transpose}\,\diag(m_1,\dots,m_n)\,U\;,
\end{align}  
\end{subequations}
where $U_\mathrm{L}$, $U_\mathrm{R}$ and $U$ are unitary matrices. 
The $m_i$ are the singular values, and their absolute values, $\lvert m_i\rvert$, are the physical masses up to contributions from the K\"ahler metric. 
In either case one has
\begin{equation}
  \lvert\det M\rvert= \lvert m_1\,\cdots\, m_n\rvert\;.
\end{equation}

Key insights derive from the fact that the determinant of mass matrix $M(\tau)$ in \eqref{eq:superpotential_mass} transforms as a 1-dimensional \ac{VVMF}, 
\begin{equation}
\det M(\tau)\xmapsto{~\gamma~} \det M(\gamma\tau)=(c\,\tau+d)^{\sum_{i} k_{\varphi^\ChargeC_i}+k_{\varphi_i}}\,\left(\det\rho^{\ChargeC}\,\rho\right)^*\,\det M(\tau)\;. \label{eq:detMTransform}
\end{equation}
Clearly, the sum of the modular weights, $k_{\det M}\defeq\sum_{i} k_{\varphi^\ChargeC_i}+k_{\varphi_i}$, is a positive integer.\footnote{If $k_{\det M}=0$, then $\det M(\tau)$ must be a constant, a fact which was used to solve the strong \CP problem~\cite{Feruglio:2023uof,Feruglio:2024ytl,Petcov:2024vph,Penedo:2024gtb}.}
Further, $\rho_{\det M}\defeq \det(\rho^\ChargeC\,\rho)^*$ is a 1-dimensional representation of $\Group{SL}{2,\mathds{Z}}$ and the finite modular group since $\rho^\ChargeC$ and $\rho$ are 3-dimensional representations of $\Group{SL}{2,\mathds{Z}}$ and the finite modular group.
Therefore, $\det M$ is indeed a certain 1-dimensional \ac{VVMF}. 

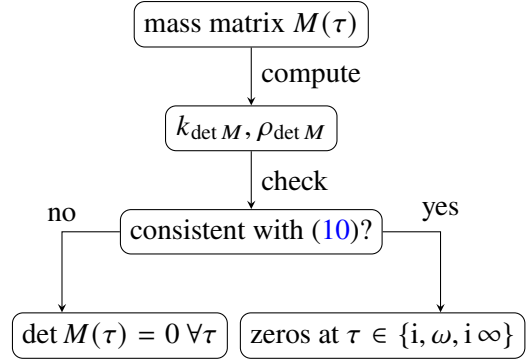
\begin{wrapfigure}[13]{r}{7cm}  
  \centering
  \begin{tikzpicture}[>=stealth]
    \path[nodes={draw,rounded corners},node distance=2em] 
      node(m){mass matrix $M(\tau)$}
      node[below=of m](det){$k_{\det M},\rho_{\det M}$}
      node[below=of det](cons){consistent with \eqref{eq:1VVMF}?}
      node[below=of cons.south west](n){$\det M(\tau)=0~\forall\tau$}
      node[below=of cons.south east](y){zeros at $\tau\in\{\ii,\omega,\ii\infty\}$};
      \draw[->] (m) -- node[right]{compute}(det);
      \draw[->] (det) --node[right]{check} (cons);
      \draw[->] (cons.west) -| node[above]{no} ([xshift=-2em]n.north);
      \draw[->] (cons.east) -| node[above]{yes} ([xshift=2em]y.north);
  \end{tikzpicture}
\caption{A simple test reveals the location of the zeros of a given modular mass matrix.}
\label{fig:DecisionTree}
\end{wrapfigure}

This has some important and immediate consequences. 
There are two options (cf.\ \Cref{fig:DecisionTree}): 
\begin{enumerate}
 \item If the modular weight $k_{\det M}$ and representation $\rho_{\det M}$ are inconsistent with \Cref{eq:1VVMF}, then $\det M$ is necessarily constantly zero. 
 Clearly this is not a viable way of explaining hierarchically small yet nonzero masses. 
 \item If the modular weight $k_{\det M}$ and representation $\rho_{\det M}$ are consistent with \Cref{eq:1VVMF}, $\det M$ will only vanish at certain points in the fundamental domain.
 As we have seen in \Cref{sec:1-dimensional_VVMF}, as long as the modular weights are not too large, these zeros can only be at $\tau\in\{\ii,\,\omega,\,\ii\infty\}$. 
\end{enumerate}

\begin{table}[th!]
    \centering
    \begin{tabular}{ccccc} 
      \toprule
      $k_{\det{M}}$ & Modular Form & Zeros & Multiplicity $(n_{\ii\infty},\, n_{\ii},\, n_{\omega})$  & Representation\\ \midrule
      1 & $\eta^2$ & $\ii\infty$ & $(1,0,0)$ & $\rep{1}_1$\\
      \midrule
      2 & $\eta^4$ & $\ii\infty$ & $(2,0,0)$ & $\rep{1}_2$\\
      \midrule    
      3 & $\eta^6$ & $\ii\infty$  & $(3,0,0)$ & $\rep{1}_3$\\
      \midrule
      4 & $E_4$ & $\omega$ & $(0,0,1)$ & $\rep{1}_0$ \\
        & $\eta^{8}$ & $\ii\infty$ & $(4,0,0)$& $\rep{1}_4$\\
      \midrule
      5 & $\eta^{2}\,E_4$ & $\omega$ and $\ii\infty$  & $(1,0,1)$ & $\rep{1}_1$ \\
        & $\eta^{10}$ & $\ii\infty$  & $(5,0,0)$ & $\rep{1}_5$ \\
      \midrule
      6 & $E_6$ & $\ii$  & $(0,1,0)$ &  $\rep{1}_0$\\
        & $\eta^4\,E_4$ & $\omega$ and $\ii\infty$  & $(2,0,1)$ & $\rep{1}_2$\\
        & $\eta^{12}$ & $\ii\infty$  & $(6,0,0)$ & $\rep{1}_6$\\
      \midrule
      7 & $\eta^{6}\,E_4$ & $\omega$ and $\ii\infty$  & $(3,0,1)$ & $\rep{1}_3$ \\
        & $\eta^{2}\,E_6$ & $\ii$ and $\ii\infty$  & $(1,1,0)$ & $\rep{1}_1$ \\
        & $\eta^{14}$ & $\ii\infty$  & $(7,0,0)$ & $\rep{1}_7$ \\
      \midrule
      8 & $E_4^2$ & $\omega$  & $(0,0,2)$ & $\rep{1}_0$\\
        & $\eta^{4}\,E_6$ & $\ii$ and $\ii\infty$  & $(2,1,0)$ & $\rep{1}_2$ \\
        & $\eta^{8}\,E_4$ & $\omega$ and $\ii\infty$  & $(4,0,1)$ & $\rep{1}_4$ \\
        & $\eta^{16}$ & $\ii\infty$ & $(8,0,0)$ & $\rep{1}_8$\\
     \midrule
      9 & $\eta^{10}\,E_4$ & $\omega$ and $\ii\infty$  & $(5,0,1)$& $\rep{1}_5$ \\    
        & $\eta^{6}\,E_6$ & $\ii$ and $\ii\infty$  & $(3,1,0)$& $\rep{1}_3$ \\  
        & $\eta^{2}\,E^2_4$ & $\omega$ and $\ii\infty$  & $(1,0,2)$& $\rep{1}_1$ \\  
        & $\eta^{18}$ & $\ii\infty$  & $(9,0,0)$ & $\rep{1}_9$ \\     
      \midrule
      10 & $E_4\,E_6$ & $\omega$ and $\ii$  & $(0,1,1)$ & $\rep{1}_0$\\
        & $\eta^{4}\,E_4^2$ & $\omega$ and $\ii\infty$  & $(2,0,2)$& $\rep{1}_2$\\
        & $\eta^{8}\,E_6$ & $\ii$ and $\ii\infty$  & $(4,1,0)$ & $\rep{1}_4$\\
        & $\eta^{12}\,E_4$ & $\omega$ and $\ii\infty$  & $(6,0,1)$& $\rep{1}_6$\\ 
        & $\eta^{20}$ & $\ii\infty$  & $(10,0,0)$ & $\rep{1}_{10}$\\
      \midrule
      11 & $\eta^{2}\,E_4\, E_6$ & $\omega$, $\ii$ and $\ii\infty$  & $(1,1,1)$ & $\rep{1}_1$\\
        & $\eta^{6}\,E_4^2$ & $\omega$ and $\ii\infty$  & $(3,0,2)$ & $\rep{1}_3$\\
        & $\eta^{10}\,E_6$ & $\ii$ and $\ii\infty$  & $(5,1,0)$ & $\rep{1}_5$\\
        & $\eta^{14}\,E_4$ & $\omega$ and $\ii\infty$  & $(7,0,1)$ & $\rep{1}_7$\\
        & $\eta^{22}$ & $\ii\infty$  & $(11,0,0)$ & $\rep{1}_{11}$\\
      \bottomrule
    \end{tabular}
  \caption{Possible modular forms that the determinants of the mass matrices may be proportional to. 
  We display the corresponding zeros and the multiplicities $(n_{\ii\infty},\, n_{\ii},\, n_{\omega},\, \sum n_{\tau_0})$ as defined in \Cref{eq:ValenceFormula}. Note that $\sum n_{\tau_0}$ is zero for all cases in the table, as the weights are less than 12.}
  \label{tab:detModularForm}
\end{table}

The arguably most plausible way to obtain hierarchies is when $\tau$ is very close to a value at which the determinant vanishes. 
$\det M$ can be expanded according to \Cref{eq:asymptotic} near a zero $\tau_0$. 
We naturally expect a similar expansion for each singular value $m_1$, $m_2$ or $m_3$, so that the mass ratios scale as
\begin{equation}\label{eq:GeneralHierarchy}
m_1\relative m_2\relative m_3 \approx c_1\,u^{n_1}\relative c_2\,u^{n_2}\relative c_3\,u^{n_3}\;,
\end{equation}
where $u$ is defined in \eqref{eq:asymptotic}. 
The hierarchy is then expected to arise from the smallness of $u$. 
This is the case which we will mainly discuss in this and the following section. 
Alternatively, the hierarchy may also originate from the coefficients $c_{1,2,3}$, which are generally determined by the model. 
We will comment on this possibility in more detail in \Cref{sec:coeff}.

Notice now that $\det M \approx c_1\,c_2\,c_3\, u^{n_1+n_2+n_3}$.
Therefore, for a given assignment of modular weights and representations in a model, we can see that the zeros $\tau_0$ and values $n_1+n_2+n_3$ can be deduced directly without the need for a detailed analysis. 
In turn, by surveying all possible 1-dimensional \acp{VVMF}, we can classify the regions of moduli space that can naturally produce the mass hierarchy and determine the overall size of $\det M$ as $u^{n_1+n_2+n_3}$.

We list the possible candidate modular forms for the determinant of the mass matrix along with their zeros for weights less or equal to 11 in \Cref{tab:detModularForm}. 
As one can see, for determinants with weights of less than 12, achieving a hierarchical fermion mass from the expansion of $\tau$ requires that $\vev{\tau}$ be close to one of the critical points.
For the weights under consideration, zeros can only occur at the critical points, as $\sum n_{\tau_0} = 0$.

On the other hand, for weights larger than 11, one may achieve zeros at points other than the critical points by forming a linear combination of Eisenstein series. 
For example, at weight 12, the determinant can take the form $a\, E_4^3 + b\, E_6^2$, where $a$ and $b$ are generally complex numbers. 
When both $a$ and $b$ are nonzero, the determinant may have zeros at points distinct from the critical points.

However, the zeros away from the critical points appear to be less relevant for fermion masses. 
In a realistic quark or charged lepton model, one would expect at least a hierarchy of $1\relative u\relative u^2$, leading to a determinant proportional to $u^3$, meaning $\tau$ must be close to a zero of at least order 3. 
If zeros occur at a point other than the critical points, the left-hand side of \Cref{eq:ValenceFormula} must contribute $n_{\tau_0} = 3$ at that zero. 
Thus, if $\tau$ acquires a \ac{VEV} at such a point, the determinant's weight must be at least 36 to ensure hierarchical masses. 
Such large weights are unusual in bottom-up models, and appear to be prohibited in top-down constructions.

Let us recap the key take-away messages from our discussion so far:
\begin{enumerate}
 \item As long as $\det M$ is a nontrivial \ac{VVMF}, it has zeros within the fundamental domain.
 \item For $k_{\det M} = \sum^{3}_{i=1} k_{\varphi^\ChargeC_i} + k_{\varphi_i} < 12$,
  \begin{enumerate}
    \item the zeros can only be at $\ii$, $\omega$, or $\ii\infty$, i.e.\ the critical points,
    \item only the point $\ii\infty$ can lead to hierarchies $u^{n>3}$, and
 \item the hierarchy is bounded from above by $k_{\det M}$, i.e.\ the power $n_1+n_2+n_3$ of the hierarchy cannot exceed $k_{\det M}$.\label{item:det_max_hierarchy}
  \end{enumerate} 
 \item For $k_{\det M} \geq 12$, zeros can occur at points other than the critical points, but the determinant's weight must be at least 36 to ensure hierarchical masses.
\end{enumerate}
Before we conclude this section, let us illustrate these points with a few examples.

\subsection{Example: Feruglio model}
\label{subsec:ExampleFeruglioModel}

A classical example in modular flavor models is the first model presented in~\cite{Feruglio:2017spp}, where neutrino masses are generated through the Weinberg operator. 
In this model, the lepton doublet $L$ transforms as a triplet under the finite modular group $A_4$, and the Higgs doublet $H_u$ is a singlet under $A_4$. 
The relevant superpotential and weight assignments are
\begin{equation}
  \mathscr{W}_\nu = \frac{1}{2\Lambda}\, (H_u\, L)^\transpose\, Y_\nu(\tau)\, (H_u\, L)\;,\quad
  \text{where } k_{H_u} = 0\text{ and } k_{L} = 1\;.\label{eq:A4Superpotential}
\end{equation}
Here, $k_{H_u}$ and $k_{L}$ are the weights of the Higgs doublet $H_u$ and the lepton doublet $L$, respectively.

From the weight assignment, one can check that the determinant must have weight 6. 
In \Cref{tab:detModularForm}, there are three options at weight 6: $\rep{1}_0$, $\rep{1}_2$, and $\rep{1}_6$.
However, it is well known that $A_4$ does not have $\rep{1}_2$ or $\rep{1}_6$ representations. 
Instead, $A_4$ has the following three singlets. 
These are $\rep{1}_0$, the trivial singlet, which often gets referred to as $\rep{1}$, as well as $\rep{1}_4$ and $\rep{1}_8$, which often get denoted as $\rep{1'}$ and $\rep{1''}$, respectively.
Since the only possible weight-6 representation is $\rep{1}_0$, we can conclude that any nontrivial determinant has to be proportional to the Eisenstein series $E_6$.
This implies that the only possible hierarchy is $1\relative 1\relative u$, suggesting an inverted ordering for neutrino masses, with the best-fit value of $\tau$ around $\tau=\ii$.

Let us verify this explicitly. 
The mass matrix for this model is given by
\begin{equation}
M_\nu = \frac{v_u^2}{2\Lambda}
\begin{pmatrix}
     2 Y_{\rep{3},1}^{(2)} & -Y_{\rep{3},3}^{(2)} & -Y_{\rep{3},2}^{(2)} \\
     -Y_{\rep{3},3}^{(2)} & 2 Y_{\rep{3},2}^{(2)} & -Y_{\rep{3},1}^{(2)} \\
     -Y_{\rep{3},2}^{(2)} & -Y_{\rep{3},1}^{(2)} & 2 Y_{\rep{3},3}^{(2)} \\
\end{pmatrix}\;,\quad
\label{eq:ExampleFeruglioModelMassmatrix}
\end{equation}
where $v_u$ is the \ac{VEV} of the $u$-type Higgs $H_u$, and $Y_{\rep{3},i}^{(2)}$ is the $i$\textsuperscript{th} component of the modular form in the representation $\rep{3}$ of weight 2 for $A_4$. Its $q$-expansion is given by
\begin{equation}
  Y_{\rep{3}}^{(2)} =
\begin{pmatrix}
  1 + 12\, q + 36\, q^2 + 12\, q^3 + 84\, q^4 + 72\, q^5 + \dots \\
  -6\, q^{\frac{1}{3}}\,\bigl(1 + 7\, q + 8\, q^2 + 18\, q^3 + 14\, q^4 +\dots\bigr)\\
  -18\, q^{\frac{2}{3}}\,\bigl(1 + 2\, q + 5\, q^2 + 4\, q^3 + 8\, q^4 +\dots\bigr)\\
\end{pmatrix}\;,\label{eq:A4Y3k2}
\end{equation}
where $q \equiv \ee^{2\pi\ii \tau}$ as before.
At first sight, this seems to suggest that the neutrino mass matrix may exhibit a hierarchy $1\relative u \relative u^2$ as $q\to0$.
However, the determinant is in fact given by
\begin{align}
 \det M_\nu = -2\frac{v_u^6}{(2\,\Lambda)^3} E_6(\tau) &= -2\frac{v_u^6}{(2\,\Lambda)^3} \bigl(1 - 504\, q - 16632\, q^2 - 122976\, q^3 + \dots\bigr) \nonumber \\
  &= -2 \frac{v_u^6}{(2\,\Lambda)^3}\left(\frac{9\,\mathnormal{\Gamma}\left(1/4\right)^{16}}{(2\,\pi)^{11}}\,u - \frac{54\,\mathnormal{\Gamma}\left(1/4\right)^{16}}{(2\,\pi)^{11}}\,u^2  
+\dots\right)\;,
\end{align}
where $u = \frac{\tau-\ii}{\tau+\ii}$. 
$\mathnormal{\Gamma}$ denotes the Gamma-function, and should not be confused with the congruence subgroup $\Gamma(N)$.
This agrees with our previous analysis, showing that the determinant has a zero of order 1 at $\tau=\ii$, and approaches a constant as $\tau\to\ii\infty$. 
Indeed, the best-fit point found in \cite{Feruglio:2017spp} is $\tau = 0.0111 + 0.9946 \ii$, i.e.\ close to $\ii$, with $\lvert u\rvert \sim 0.0062$.
The model predicts an inverted ordering for neutrino masses, with a hierarchy of $2.24564 \relative 2.24564 \relative 5.28092\, u$.

Let us point out that these findings are somewhat complementary to the conclusions one can draw from building holomorphic modular invariant combinations of the components of $Y_{\rep{3}}^{(2)}$ \cite{Chen:2024otk}.
The latter allows us to make statements independently of $\tau$ whereas in our analysis we use properties of the determinant to conclude that $\tau$ has to be close to $\ii$. 
Neither of these conclusions requires detailed scans over the fundamental domain of $\tau$. 
Rather, they are generic features of a given modular weight and representation assignment.

\subsection{Example: \texorpdfstring{$3\times3$}{3x3} mass matrices whose determinants have weight \texorpdfstring{$3$}{3}}
\label{subsec:Example3x3Weight3}

It is instructive to look at a slightly different example.
We specifically focus on the the charged lepton masses in the \ac{SM}. 
The only assumption that we make about this mass matrix is that its determinant has weight 3.
This is realized in many examples. For instance, in a global \ac{SUSY} model we could respectively assign modular weights of 1 and 0 to the lepton doublet $L$ and charged lepton singlet $E^\ChargeC$, while assuming the Higgs doublet $H_d$ has weight 0.
It is, however, important to note that in most of the explicit \ac{SM}-like compactifications of closed strings the leading order Yukawa couplings between twisted states also have weight 1, thus leading to determinants of weight 3 for 3 generations. 

Without further information on the model, one can conclude from \Cref{tab:detModularForm} that the determinant of the mass matrix must be proportional to $\eta^6$.
If one or more mass eigenvalues are zero, then the determinant is simply zero for all $\tau$.
However, if all mass eigenvalues are to be nonzero, the determinant is uniquely determined. The zeros of the determinant occur at $\tau=\ii\infty$, implying that hierarchical fermion masses can only be achieved near the critical point $\ii\infty$. 
Moreover, we also know that the determinant transforms as $\rep{1}_3$, so the only possible modular flavor symmetries are those which have this representation.

Let us verify this explicitly by choosing a finite modular group and assigning representations to the matter fields.  
We choose the finite modular group $S_4' =\Group{SL}{2,\mathds{Z}_4}$ and adopt the representations, \ac{CG} coefficients, and modular forms from references~\cite{Liu:2020msy,Liu:2020akv}.
We assign the three generations of lepton doublets, $L$, to transform as the representation $\rep{\hat 3'}$, while the three generations of charged lepton singlets, $E^\ChargeC$, to transform as $\rep{3}$, and the Higgs doublet $H_d$ to transform as the trivial singlet $\rep{1}$.
The relevant superpotential is then
\begin{equation}
  \mathscr{W}_e = \alpha\, (E^{\ChargeC})^{\transpose}\, Y_e(\tau)\,(H_d\, L)\;,\label{eq:eSuperPotential}
\end{equation}
where $\alpha$ is a complex coefficient.  
Using the \ac{CG} coefficients, we find that the mass matrix for this model is given by
\begin{equation}
M_e = v_d\, \alpha\,
  \begin{pmatrix}
       0 &  Y_{\rep{\hat 3'},2}^{(1)} &  -Y_{\rep{\hat 3'},3}^{(1)} \\
        Y_{\rep{\hat 3'},2}^{(1)} &  Y_{\rep{\hat 3'},1}^{(1)} & 0 \\
        -Y_{\rep{\hat 3'},3}^{(1)} & 0 &  -Y_{\rep{\hat 3'},1}^{(1)} \\
  \end{pmatrix}\;,
  \label{eq:Example3x3Weight3Massmatrix}
\end{equation}
where $v_d$ is the \ac{VEV} of the $d$-type Higgs $H_d$, and
$Y_{\rep{\hat 3'},i}^{(1)}$ is the $i$\textsuperscript{th} component of the modular form in the representation $\rep{\hat 3'}$ of weight 1 in $S_4'$. 
The $q$-expansion of this modular form is given by \cite{Liu:2020msy}
\begin{equation}
	\label{eq:q-expansionOfY3hat}
  Y_{\rep{\hat 3'}}^{(1)} (\tau) =
  \begin{pmatrix}
       -2 \sqrt{2}\, q^{\frac{1}{4}}\, \bigl(1 + 2\, q + q^2 + 2\,q^3 + 2\, q^4 + 3\, q^6 + \dots\bigr) \\
       -4\, q^{\frac{1}{2}}\, \bigl(1 + 2\, q^2 + q^4 + 2\, q^6 + \dots\bigr) \\
       1 + 4\, q + 4\, q^2 + 4\, q^4 + 8\, q^5 + \dots\\
  \end{pmatrix}\;.
\end{equation}
It is then straightforward to verify that
\begin{equation}
  \det M_e = 2 \sqrt{2}\,v_d^3\, \alpha^3\, \eta^6(\tau) = 2 \sqrt{2}\, v_d^3\,\alpha^3\, \bigl( u^3 - 6 \, u^{15} + \dots\bigr) \;,
\end{equation}
where $ u = q^{1/12}$. 
This model gives a hierarchy of 
\begin{equation}\label{eq:HierarchyOfModel3.3}
m_1\relative m_2\relative  m_3 \approx	1\relative 1\relative 2\sqrt{2}\,  u^3\;.
\end{equation}
This result agrees with our previous analysis, showing that the determinant has a zero of order 3 at $\tau=\ii \infty$.

This example illustrates that, given the weight assignments, the zeros of the determinant of the mass matrix can be predicted at the critical points without relying on specific model details. 
As expected, to achieve a hierarchical mass matrix, the model must have $\vev{\tau}$ near the corresponding zeros.

Let us note that, although in our example the determinant has weight 3 because it is the determinant of a $3\times3$ mass matrix with entries of weight 1. 
Matter fields may have different modular weights, and thus the entries of the mass matrix can be modular forms of different weights. Our conclusions also apply in this case.

\subsection{Example: vanishing determinant}
\label{subsec:Example0Det}

Finally, let us consider an example in which the weight and representation of the determinant is not consistent with \Cref{eq:1VVMF}, which immediately 
allows us to conclude that the determinant vanishes identically.

To keep the example simple, we extend $A_4$ to $T'$, the double cover of $A_4$. 
We adopt the representations and tensor product results from \cite{Liu:2019khw}.
It is important to note that $T'$ and $A_4$ share the same representations and \acp{VVMF} for the singlets and triplet. 
Therefore, we will again use the triplet \ac{VVMF} in \Cref{eq:A4Y3k2}.

Here we consider a two-generation example, which alternatively also may be viewed as an embedding in a $\rep{2} \oplus \rep{1}$ reducible representation. 
Focusing on the neutrino sector, we again use the superpotential and weight assignments from \Cref{eq:A4Superpotential}. 
That is, the neutrino mass is generated through the Weinberg operator, with the lepton doublet $L$ having weight 1 and the Higgs weight 0. 
We further assign $L$ to transform as one of the doublets, say $\rep{2'}$. 
This leads to a mass matrix of the form
\begin{equation}\label{eq:Mass_matrix_vanishing_determinant}
  M_\nu = \frac{v_u^2}{2\,\Lambda}
  \begin{pmatrix}
       -\sqrt{2}\, Y_{\rep{3},3}^{(2)} & Y_{\rep{3},2}^{(2)} \\
       Y_{\rep{3},2}^{(2)} & \sqrt{2}\, Y_{\rep{3},1}^{(2)} \\
  \end{pmatrix}\;,\quad
\end{equation}
where $v_u$ is the \ac{VEV} of the $u$-type Higgs $H_u$, and $Y_{\rep{3},i}^{(2)}$ is the $i$\textsuperscript{th} component of the modular form defined in \Cref{eq:A4Y3k2}.

Now, the determinant of the mass matrix is given by
\begin{equation}
  \det M_\nu = -\frac{v_u^4}{4\,\Lambda^2}\left(2 \, Y_{\rep{3},3}^{(2)}\, Y_{\rep{3},1}^{(2)} + \left(Y_{\rep{3},2}^{(2)}\right)^2\right)\;.
\end{equation}
Since this is a $2\times2$ matrix the entries of which have weight 2, $\det M_\nu$ has modular weight 4.
It can be further verified that $\det M_\nu$ transforms as $\rep{1}_8$.
Indeed, a quick cross-check yields
\begin{equation}
  \left(Y_{\rep{3}}^{(2)}\, Y_{\rep{3}}^{(2)}\right)_{\rep{1}_8} = 2 \, Y_{\rep{3},3}^{(2)} \, Y_{\rep{3},1}^{(2)} + \left(Y_{\rep{3},2}^{(2)}\right)^2 \;.
\end{equation}
So $\det M_\nu$ is a 1-dimensional \ac{VVMF} of weight 4, transforming as $\rep{1}_8$.
However, as can be seen from \Cref{eq:1VVMF}, there is no \ac{VVMF} of weight 4 transforming as $\rep{1}_8$. 
This can also be verified by inspecting \Cref{tab:detModularForm}. 
Therefore, the determinant of the mass matrix \eqref{eq:Mass_matrix_vanishing_determinant} must vanish. 

Notice that this result can also be obtained from the algebraic relation \cite{Feruglio:2017spp}
\begin{equation}
  2 \,Y_{\rep{3},3}^{(2)} \,Y_{\rep{3},1}^{(2)} + \left(Y_{\rep{3},2}^{(2)}\right)^2 = 0 \;. 
\end{equation}
This follows from the fact that $Y_{\rep{3}}^{(2)}$ can be constructed from the tensor product of the lowest weight modular forms $Y_{\rep{2}}^{(1)}$ \cite{Liu:2019khw}, and which has also been used in \cite{Chen:2024otk}.

\subsection{Conclusions from Section \thesection}

From the above examples, we see how zeros of determinants give us insight into models with hierarchical masses. 
We close this section with a few comments.
First, these results imply important constraints on the choice of modular flavor symmetries. We have argued that pronounced mass hierarchies arise only at $\tau= \ii\infty$. 
As one can see from \Cref{tab:detModularForm}, for weights smaller than 12 this implies that the determinant of the mass matrix has to be a nontrivial 1-dimensional \ac{VVMF}. 
This requirement rules out all groups which do not have nontrivial 1-dimensional representations. 
It can be shown that the latter coincide with the so-called perfect groups \cite[Appendix~A]{Chen:2015aba}, which include important examples such as $\Gamma_5\cong A_5$.

It may be possible to avoid this conclusion by introducing additional flavons into the model. 
In such scenarios, the mass matrix now depends on both $\tau$ and the flavons, $M(\tau,\chi_n)$, where $\chi_n$ represents the flavons. 
When the flavons acquire \acp{VEV}, the mass matrix becomes $M(\tau,\vev{\chi_a})$, where $\vev{\chi_a}$ is the \ac{VEV} of the flavons, which no longer transforms. This breaks $\Group{SL}{2,\mathds{Z}}$, and one needs additional information beyond \Cref{tab:detModularForm} to determine the hierarchy~\cite{King:2020qaj,Ding:2025mar}.

Altogether we thus see that the critical points $\ii$, $\omega$, and $\ii \infty$ play a key role as they, in the absence of accidental cancellations, indicate where zeros of the determinant of the mass matrix occur. 
This means that mass hierarchies arise in the neighborhoods of these critical points. 
In fact, as we will show next, the mass hierarchy structure, $ u^{n_1}\relative  u^{n_2}\relative  u^{n_3}$, can be predicted at these points without relying on specific model details.

\section{Modular symmetries near critical points}
\label{sec:critical}

\subsection{Universal near-critical behavior}
\label{sec:near-critical-behavior}

\begin{wraptable}[10]{r}{8.5cm}
  \centering\vspace*{-1ex}
  \begin{tabular}{cccc} \toprule
    Fixed Point $\tau_0$    & $\ii$ & $\omega = \ee^{2\pi\ii/3}$ & $\ii\infty$\\ \midrule
    Stabilizer $G_0$      & $\mathds{Z}_4^{\GeneratorS}$ & $\mathds{Z}_3^{\GeneratorS\,\GeneratorT}\times \mathds{Z}_2^{\GeneratorS^2}$               & $\mathds{Z}^{\GeneratorT}\times \mathds{Z}_2^{\GeneratorS^2}$  \\ 
    Order of the group         & 4   & 6                              & $\infty$ \\ \bottomrule
  \end{tabular}
\caption{Fixed points of the modular group, and the corresponding stabilizers which are subgroups of $\Group{SL}{2,\mathds{Z}}$. Note that $\mathds{Z}_2^{\GeneratorS^2}$ stabilizes generic points in moduli space, and acts on matter fields like a parity.}
\label{tab:listOfFixedPoints}
\end{wraptable}
In this section, we study the behavior of the theory near the points $\ii$, $\omega$, and $\ii \infty$, known as the critical or fixed points of the modular flavor symmetry. 
From this point onward, we will use the notation $\tau_0$ to denote a fixed point of the modular group $\Group{SL}{2,\mathds{Z}}$. 
These points are referred to as fixed or critical points because each remains invariant under a stabilizer group of the modular group $\Group{SL}{2,\mathds{Z}}$. 

Generally, modular flavor symmetries are nonlinearly realized, as shown in \Cref{eq:modulartransf}, and at a generic point in $\mathcal{F}$, the symmetry is completely broken. 
However, if $\tau$ acquires a \ac{VEV} at a fixed point $\tau_0$, a linearly realized residual symmetry can remain. This is because $\tau_0$ is left unchanged under the stabilizer group $G_0$, which is an abelian subgroup of $\Group{SL}{2,\mathds{Z}}$. 
We summarize the fixed points and their corresponding residual symmetries in \Cref{tab:listOfFixedPoints}.

\begin{figure}[htb]
\centering\subcaptionbox{The Poincar\'e disk around $\ii$.\label{fig:PoinareDiskAroundI}}
{\includegraphics[width=0.46\textwidth]{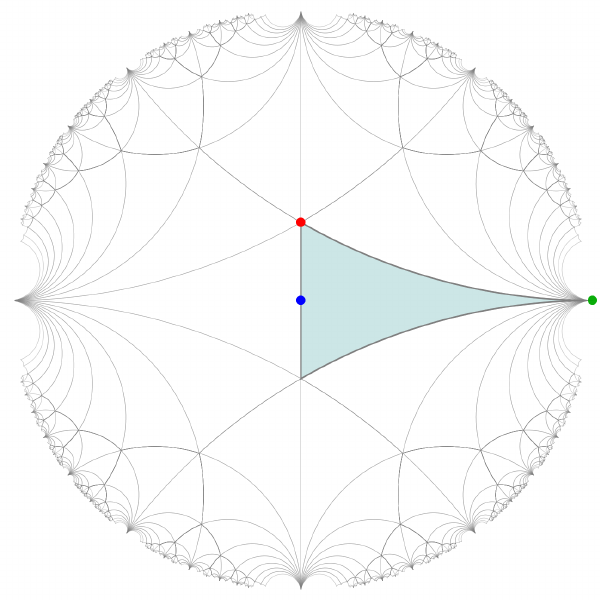}} \qquad
 \subcaptionbox{The Poincar\'e disk around $\omega$.\label{fig:PoinareDiskAroundOmega}}
{\includegraphics[width=0.46\textwidth]{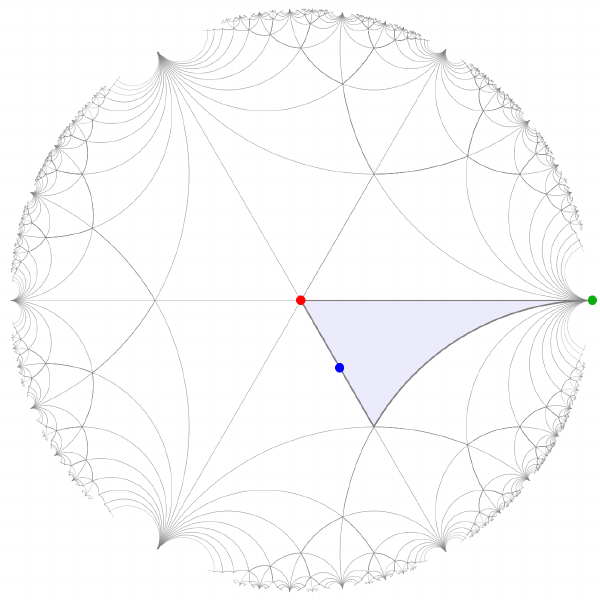}} 
\caption{The Poincar\'e disks, along with the corresponding $\Group{SL}{2,\mathds{Z}}$ fundamental domain. The blue dots, red dots, and green dots represent the images of the fixed points $\ii$, $\omega$ and $\ii\infty$, respectively. 
The fundamental domain is represented by the shaded hyperbolic triangle within the disk.}
\label{fig:PoinareDisk}
\end{figure}

It is usually more straightforward to analyze a symmetry when it is linearly realized. 
Given a stabilizer group $G_0$ with its corresponding fixed point $\tau_0$, the originally nonlinear modular transformations in~\Cref{eq:modulartransf} can be linearized within the given stabilizer group.\footnote{Notice the linearized symmetry is part of the original modular symmetry $\Group{SL}{2,\mathds{Z}}$, which is still nonlinearly realized overall.}
To achieve this, motivated by \Cref{eq:asymptotic}, we perform the field redefinitions \cite{Bruinier:2008xxx,Feruglio:2021dte,Novichkov:2021evw}\footnote{Note the difference between the definition of $\varepsilon$ and the definition of $u$ in \Cref{eq:asymptotic} when $\tau_0 = \ii\infty$. $u$ may be regarded as an expansion parameter defined for level $N = 12$. The reason is that the 1-dimensional \acp{VVMF} are (scalar) modular forms of level $N = 12$, and the corresponding finite modular group is the Abelian group $\mathds{Z}_{12}$. $\varepsilon$ defined in \eqref{eq:fieldRedef_b} is a generalization for any general level $N$. It is related to $u$ from \eqref{eq:asymptotic} via $u = \varepsilon^{N/12}$. }  
\begin{subequations}\label{eq:fieldRedef}
\begin{align}
 &\left.\begin{aligned} 
  \varepsilon &\defeq \eqmathbox[fr][l]{\frac{\tau - \tau_0}{\tau - \bar{\tau}_0}} \\
  \Phi&\defeq\eqmathbox[fr][l]{(1-\varepsilon)^{k_\varphi}\,\varphi}
 \end{aligned}\right\} \text{ for } \tau_0 = \ii ~\text{or}~\omega\;, \label{eq:fieldRedef_a}\\
 &\left.\begin{aligned} \varepsilon &\defeq\eqmathbox[fr][l]{\ee^{2\pi\ii\tau/N}}\\
  \Phi&\defeq\eqmathbox[fr][l]{\varphi}
  \end{aligned}\right\} \text{ for } \tau_0= \ii \infty\;,\label{eq:fieldRedef_b}
\end{align}
\end{subequations}
where $N$ is the order of $\rho(\GeneratorT)$, which is also known as the level of the finite modular group. 
For illustration, we present the moduli space of the new modulus $\varepsilon$ in \Cref{fig:PoinareDisk}. 
Specifically, these moduli spaces are images of the upper half plane $\mathcal{H}$ of $\tau$ via the Cayley mapping given in \Cref{eq:fieldRedef_a}. 
They are commonly referred to as Poincar\'e disks. 
Among them, \Cref{fig:PoinareDiskAroundI} and \Cref{fig:PoinareDiskAroundOmega} are respectively unit disks centered at $\ii$ and $\omega$. 
The highlighted regions correspond to the original fundamental domain of $\Group{SL}{2,\mathds{Z}}$, as shown in \Cref{fig:Critical_points}.

In the field basis of \Cref{eq:fieldRedef}, modular transformations by an element of the stabilizer group $\gamma_0\in G_0$ take the form
\begin{subequations}\label{eq:LinearizedTrafos}
\begin{align}
 \varepsilon&\xmapsto{~\gamma_0~}{\ee^{\ii\theta_0}\, \varepsilon}\;,\\ 
 \Phi&\xmapsto{~\gamma_0~}j_0^{k_\varphi}\,\rho(\gamma_0)\,\Phi\eqdef\Omega_\varphi(\gamma_0)\,\Phi\;, \label{eq:36b}
\end{align}
\end{subequations}
where $\theta_0$ and $j_0$ can be found in \Cref{tab:LinearizedTrafoPhases}.
$\Omega_\varphi$ is often called the weighted representation, which still belongs to the representation of $\Group{SL}{2,\mathds{Z}}$, and therefore of the subgroup $G_0$.

\begin{wraptable}[8]{r}{6cm}
  \centering$\begin{array}{*5c}
    \toprule 
    \gamma_0 & \GeneratorS & \GeneratorS\,\GeneratorT & \GeneratorT & \GeneratorS^2\\
    \midrule 
    \theta_0 & \pi & 4\,\pi/3 & 2\,\pi/N & 0\\
    j_0 & \ii & \omega & 1 & -1\\
    \bottomrule
   \end{array}$  
  \caption{$\theta_0$ and $j_0$ for \Cref{eq:LinearizedTrafos}.}
  \label{tab:LinearizedTrafoPhases} 
\end{wraptable}

Since we work with representations with finite image, this means that we can switch to a basis in which $\Omega_\varphi$ is diagonal for the given stabilizer group,
\begin{equation}
  \Omega_{\varphi}(\gamma_0) = \diag \bigl(\ee^{\ii\alpha_1},\, \ee^{\ii\alpha_2}, \ee^{\ii\alpha_3}\bigr)\;.
  \label{eq:generalOmegaGamma0}
\end{equation}
This is equivalent to providing the representation decomposition of the matter fields under the stabilizers, which is a direct sum of 1-dimensional irreducible representations of stabilizers.

For the redefined modulus $\varepsilon$, the phase factor $\ee^{\ii\theta_0}$ also can be thought of as a 1-dimensional irreducible representation of the stabilizer $G_0$. 
This shows that the emerging scheme is qualitatively similar to the \ac{FN} scenario.
That is, the linearized symmetries may be thought of as $\mathrm{U}(1)_{\textnormal{FN}}$ symmetries, or $\mathds{Z}_n$ subgroups thereof, under which the matter fields transform. 
The linearized modulus $\varepsilon$ plays the role of the traditional flavon with its charge being completely fixed as shown in~\Cref{tab:LinearizedTrafoPhases}.

We now demonstrate that, despite the vast number of finite modular groups and the numerous representations of a given dimension, there are only a few types of representation decompositions under the Abelian stabilizer group $G_0 = \mathds{Z}_4^{\GeneratorS}$ or $\mathds{Z}_{3}^{\GeneratorS\,\GeneratorT} \times \mathds{Z}_2^{\GeneratorS^2}$. \
The linearized modular flavor transformations of the matter fields can be universally classified into only a few distinct scenarios.

In terms of the redefined fields, the lepton sector contains the chiral multiplets $H_u$, $H_d$, $E^\ChargeC$ and $L$ representing the $u$- and $d$-type Higgs, charged lepton singlets and doublets, respectively. 
The effective superpotential for the lepton sector is then given by
\begin{align}
   \mathscr{W}(\varepsilon,\Phi) &= \frac{1}{2\,\Lambda_L}\,\bigl(H_u\, L\bigr)^{\transpose}\, Y_\nu(\varepsilon)\,\bigl(H_u\, L\bigr) + \bigl(E^{\ChargeC}\bigr)^{\transpose}\, Y_e(\varepsilon)\,\bigl(H_d\, L\bigr)\;,
\end{align}
where $\Lambda_L$ denotes the see-saw scale, and $Y_\nu(\varepsilon)$ and $Y_e(\varepsilon)$ are, in general, linear combinations of \acp{VVMF}.\footnote{One can make the K\"ahler potential canonical to obtain the physical mass matrix $M(\varepsilon,\bar{\varepsilon})$ from $Y(\varepsilon)$. It is worth noting that even for the most general Hermitian K\"ahler matrix $K_\Phi(\varepsilon,\bar{\varepsilon})$, the transformation \eqref{eq:massMatrixTransformationRule} remains unchanged after moving to the canonical field basis, except that all holomorphic $M(\varepsilon)$ become non-holomorphic $M(\varepsilon,\bar{\varepsilon})$, which does not affect our analysis here~\cite{Feruglio:2023mii}.}
The mass matrices obey the transformation laws
\begin{subequations}\label{eq:massMatrixTransformationRule}
  \begin{align}
   M_\nu(\varepsilon)                           
   &\xmapsto{~\gamma_0~}
   M_\nu\bigl(\ee^{\ii\theta_0}\,\varepsilon\bigr) =\Omega^*\,  M_\nu(\varepsilon)\, \Omega^\dagger\, \Omega_{\mathscr{W}}\;, \\
   M_\nu^{-1} (\varepsilon)                     
   &\xmapsto{~\gamma_0~}
   M_\nu^{-1}\bigl(\ee^{\ii\theta_0}\,\varepsilon\bigr)=\Omega \, M_\nu^{-1}(\varepsilon)\, \Omega^\transpose\, \Omega_{\mathscr{W}}^*\;,   \\
   M_e(\varepsilon)                             
   &\xmapsto{~\gamma_0~}
   M_e\bigl(\ee^{\ii\theta_0}\,\varepsilon\bigr)=\Omega_\ChargeC^*\, M_e(\varepsilon) \,\Omega^\dagger\, \Omega_{\mathscr{W}}\;, \\
  M_{\bar e e}(\varepsilon,\bar{\varepsilon})   
  &\xmapsto{~\gamma_0~}
  M_{\bar e e}\bigl(\ee^{\ii\theta_0}\,\varepsilon,\ee^{-\ii\theta_0}\,\bar{\varepsilon}\bigr)=
  \Omega\, M_{\bar e e}(\varepsilon,\bar{\varepsilon})\,\Omega^\dagger\;, 
  \end{align}
\end{subequations}
where
\begin{equation}
\label{eq:omegaDefinition}
      \Omega\defeq\Omega_{H_u}\, \Omega_{L}\;, \quad 
      \Omega_\ChargeC \defeq \Omega_{H_u}^*\,\Omega_{H_d}\,\Omega_{E^\ChargeC}\quad\textnormal{and}\quad 
      M_{\bar e e}\defeq M_e^\dagger\, M_e \;.
\end{equation}
If the neutrino mass arises from the see-saw mechanism, $M_\nu$ could become singular at $\tau_0$, i.e.\ $\lim_{\tau\to\tau_0}\det(M_\nu)=\infty$.
Therefore, it is useful to also discuss the transformation behavior of $M_\nu^{-1}$. 
According to \Cref{tab:LinearizedTrafoPhases}, $\Omega_{\mathscr{W}}$ is an overall phase coming from the automorphy factor for nontrivial modular weight $k_{\mathscr{W}}$ as well as the representation ``matrix'' $\rho_{\mathscr{W}}$ of the superpotential. 
In the case of trivial combination, such as $k_{\mathscr{W}}=0$ and $\rho_{\mathscr{W}}=\rep{1}$, this will leads to $\Omega_{\mathscr{W}} = 1$. 
Nontrivial $\Omega_{\mathscr{W}}$ do not lead to qualitatively different results, as explained in~\Cref{app:classification}.

Notice that, as shown in \Cref{eq:massMatrixTransformationRule}, for Majorana neutrinos, both flavor mixing and neutrino mass can be inferred solely from $\Omega$, i.e.\ the transformation properties of the left-handed leptons $L$.
On the other hand, for Dirac fermions (whether charged leptons or Dirac-type neutrinos), the mass matrices also depend on $\Omega_\ChargeC$, which is determined by the right-handed fields. However, the Hermitean matrix $M_{\bar e e}$ depends only on the left-handed lepton $L$ assignment.

The transformation rules \eqref{eq:massMatrixTransformationRule} completely determine the form of the mass matrix in the neighborhood of a critical point. 
That is, we can expand the entries of a given mass matrix near a given critical value, at which $\varepsilon = 0$, as\footnote{In the case where $M(\varepsilon,\bar \varepsilon)$ depends on both $\varepsilon$ and $\bar\varepsilon$, we expand $ M_{ij}(\varepsilon,\bar \varepsilon) = m^{(0)}_{ij} + m^{(10)}_{ij}\, \varepsilon + m^{(01)}_{ij} \, \bar{\varepsilon}  +m^{(20)}_{ij} \, \varepsilon^2+  m^{(11)}_{ij}\, \varepsilon\, \bar \varepsilon  + m^{(02)}_{ij} \bar \varepsilon^2 + \dots \;.$}
\begin{equation}\label{eq:Expansion}
     M_{ij}(\varepsilon) = m^{(0)}_{ij} + m^{(1)}_{ij}\,  \varepsilon + m^{(2)}_{ij} \, \varepsilon^2 + \dots \;.
\end{equation}
The coefficients $m^{(k)}_{ij}$ in \Cref{eq:Expansion} come from the modular forms $Y(\varepsilon)$ themselves, and can be calculated from the $q$-expansion of $Y(\varepsilon)$. 
That is, in a concrete model, we can compute the coefficients $m^{(k)}_{ij}$ explicitly. 
Whether or not the expansion \eqref{eq:Expansion} allows us to infer the pattern of the mass matrices depends on the size of the expansion coefficients, $m^{(k)}_{ij}$.
Near the cusp $\ii\infty$, the expansion coefficients actually correspond to the Fourier coefficients of the \ac{VVMF}. 
The estimation of its upper bound has always been an important research topic in mathematics. 
Currently, for the nonvanishing \ac{VVMF} in representation with a finite image focused on in this work, the general result of the upper bound of its Fourier coefficients is~\cite{knopp2003vector,bajpai2023growth}
\begin{equation}\label{eq:bounds}
m^{(n)}_{ij} \approx \mathcal{O}(n^{k+2\alpha})\;,
\end{equation}
where $\alpha$ is a constant depending on the representation $\rho$ of the \ac{VVMF}, and $k$ denotes the modular weight. 
Moreover, for the cusp form components, we have $m^{(n)}_{ij} \approx \mathcal{O}(n^{k/2+\alpha})$. 
For the case of expansion near the elliptic fixed point $\ii$ or $\omega$, the expansion coefficients are also known as the Taylor expansion coefficients. 
Despite the fact that the Taylor coefficients often enjoy nontrivial relations, we are not aware of precise upper bounds on their size. 
Nevertheless, one may expect that for moderate weights the coefficients cannot be arbitrarily large. 
In \Cref{app:Estimate}, we provide a rough upper bound estimate of the Taylor coefficients and find that they have similar bounds to the Fourier coefficients as shown in \Cref{eq:bounds}.

Applying the linear transformation rule~\eqref{eq:massMatrixTransformationRule} under $G_0$ to the expansion~\eqref{eq:Expansion}, we can infer the pattern of mass matrices. 
In the neighborhood of the critical points, only the leading order terms of the expansion~\eqref{eq:Expansion} contribute significantly.
Therefore, we can expect each element of the mass matrix to exhibit power-law behavior near the critical point. 
As a consequence, the flavor observables will also exhibit power-law behavior, which we call near-critical behavior in modular flavor models.

We refer to these as the universal near-critical behaviors around the fixed points, determined by the decomposition of the matter fields' representation of $\Group{SL}{2,\mathds{Z}}$ under its Abelian stabilizer. 
We present a few examples to illustrate how \Cref{eq:massMatrixTransformationRule} determines the form of the mass matrix.

\subsection{Examples}

In \Cref{subsec:ExampleFeruglioModel,subsec:Example3x3Weight3}, we presented several simple example models. 
We now analyze their behavior around the fixed points using \Cref{eq:massMatrixTransformationRule}. 
To this end, we list the weighted representations from \Cref{eq:generalOmegaGamma0} in \Cref{tab:ExampleNearCritical}.

\begin{table}[t]
  \centering\begin{tabular}{l*{8}{c}}
    \toprule
    Example & $k_L$ & $k_{E^\ChargeC}$ & $\tau_0$     & $\gamma_0$ &  $\Omega$                        & $\Omega_\ChargeC$                        \\
    \midrule
   Section~\ref{subsec:ExampleFeruglioModel}     & $1$     & $-$         & $\ii$        & $\GeneratorS$        & $\diag (\ii,\ii,-\ii)$                            & $-$                                 \\[0.5ex]
   Section~\ref{subsec:Example3x3Weight3}     & $1$   & $0$       & $\ii \infty$ & $\GeneratorT$        & $\diag (-\ii,1,-1)$ & $\diag (-1,-\ii,\ii)$ \\
    \bottomrule
  \end{tabular}
  \caption{The weighted representations in examples of \Cref{subsec:ExampleFeruglioModel} and \Cref{subsec:Example3x3Weight3} in the $\rho(\gamma_0)$ diagonalized basis.
  }
  \label{tab:ExampleNearCritical}
\end{table}

Given the weighted representation, one can expand the mass matrix in terms of $\varepsilon$ and retain only the terms consistent with \Cref{eq:massMatrixTransformationRule} to determine the mass matrix pattern. 
For the example in \Cref{subsec:ExampleFeruglioModel}, the mass matrix pattern to leading order is
\begin{equation}
  M_\nu^{(\text{Pattern})} \sim
  \begin{pmatrix}
    m^{(1)}_{11}\,\varepsilon & m^{(1)}_{12}\,\varepsilon & m^{(0)}_{13} \\
    m^{(1)}_{21}\,\varepsilon & m^{(1)}_{22}\,\varepsilon & m^{(0)}_{23} \\
    m^{(0)}_{31} & m^{(0)}_{32} & m^{(1)}_{33}\,\varepsilon \\
  \end{pmatrix}\;,
  \label{eq:ExampleFeruglioModelMassmatrixPattern}
\end{equation}
where $m^{(k)}_{ij}$ denotes the coefficient of the $i,j$ entry at order $k$, cf.\ \Cref{eq:Expansion}, and $\varepsilon = \frac{\tau - \ii}{\tau +\ii}$ measures the departure from the critical point, cf.\ \Cref{eq:fieldRedef}.

Indeed, one can verify that the mass matrix in \Cref{eq:ExampleFeruglioModelMassmatrix} can be written as
\begin{equation}
  M_\nu \sim 
  \begin{pmatrix}
    5.28\, \varepsilon & 0 & -1.59 \\
    0 & 5.28\, \varepsilon & -1.59 \\
    -1.59 & -1.59 & 0
  \end{pmatrix}\;,
\end{equation}
after moving to the basis where $\rho(\GeneratorS)$ is diagonal, and $\varepsilon=\frac{\tau-\ii}{\tau+\ii}$. 
This agrees with the derived mass matrix pattern in~\Cref{eq:ExampleFeruglioModelMassmatrixPattern}, except that the coefficient $m^{(1)}_{12}$, $m^{(1)}_{21}$ and $m^{(1)}_{33}$ here happen to be exactly zero. 
These zeros originate from the intrinsic properties of the modular forms and the \ac{CG} coefficients in this example. 
Such effects are not captured by our near-critical analysis, and we will discuss them further in \Cref{sec:coeff}.

Similarly, for the example in \Cref{subsec:Example3x3Weight3}, using \Cref{tab:ExampleNearCritical} and \Cref{eq:massMatrixTransformationRule}, the mass matrix pattern to leading order is 
\begin{equation}
  M_e^{(\text{Pattern})}\sim
  \begin{pmatrix}
    m^{(3)}_{11}\,\varepsilon^3 & m^{(2)}_{12}\,\varepsilon^2 & m^{(0)}_{13} \\
    m^{(2)}_{21}\,\varepsilon^2 & m^{(1)}_{22}\,\varepsilon & m^{(3)}_{23}\,\varepsilon^3 \\
    m^{(0)}_{31} & m^{(3)}_{32}\,\varepsilon^3 & m^{(1)}_{33}\,\varepsilon \\
  \end{pmatrix}\;.
  \label{eq:Example3x3oModelMassmatrixPattern}
\end{equation}
Since $S_4' = \Group{SL}{2,\mathds{Z}_4}$ is a finite modular group of level 4, we take $N=4$, cf.\ our discussion around \Cref{eq:fieldRedef}.
In particular, $\varepsilon = \ee^{2\pi\ii \tau /4}$.
One can verify that this agrees with \Cref{eq:Example3x3Weight3Massmatrix} after expanding the leading term  
\begin{equation}
  M_e \sim 
  \begin{pmatrix}
    0 & -4\, \varepsilon^2 & -1 \\
    -4\, \varepsilon^2 & -2\sqrt{2}\, \varepsilon & 0 \\
    -1 & 0 & -2\sqrt{2}\, \varepsilon 
  \end{pmatrix}\;.
\end{equation}
Note that the original mass matrix is already in the basis where $\rho(\GeneratorT)$ is diagonal.
Once again, all the zero and nonzero coefficients are determined by the modular forms.

\subsection{Classification}\label{sec:classification}

In this section, we have systematically investigated all the near-critical patterns of the mass matrices near $\ii$ and $\omega$ under the $\rep{2}\oplus\rep{1}$ assignments.
The detailed classification results are presented in \Cref{app:classification}. 
A survey of the possible mass hierarchies near these two critical points, $\ii$ and $\omega$, can be found in \Cref{app:masshierarchies}. 
The 2- and 1-dimensional irreducible representations of $\Group{SL}{2,\mathds{Z}}$ with finite images are summarized in \Cref{app:1and2irreps}.
This already covers the 2- and 1-dimensional representations of the usual finite modular groups $\Gamma_N$ or $\Gamma'_N$.

Note that our results in \Cref{eq:NeutrinoMassMatrix2+1} fully contain the results that were previously obtained under the 3-dimensional irreducible representation assignments~\cite{Feruglio:2023mii}. 
In particular, we find that modulus values close to $\omega$ may also generate a mass matrix pattern similar to what one has near $\ii$.
This pattern is quite effective in explaining the experimental data of the neutrino sector, i.e.\ the neutrino masses and mixing angles. 
That is, our analysis implies that the appealing features of the near-critical points
\cite{Feruglio:2023bav} arise in a wider scope than previously appreciated, in particular also in the $\TpO$ setup.

As for the results shown in \Cref{app:masshierarchies}, it can be verified that suitable neutrino masses can be obtained near $\tau=\ii$. 
However, no charged lepton mass hierarchies emerge in this region. 
On the other hand, qualitatively realistic neutrino and charged lepton masses can be obtained near $\tau=\omega$.
However, for the relatively larger quark mass hierarchies, neither $\ii$ nor $\omega$ appear sufficient, as we have already mentioned earlier. 
Being close to $\ii\infty$ seems to be the only option for the quark sector. 
We leave the classification of the hierarchies near $\ii\infty$ for future work.

\section{Discussions}
\label{sec:Discussion}

\subsection{Preference for the fixed point at $\tau=\ii\infty$}
\label{sub:preference_for_the_fixed_point_at_ii_infty_}

Our analysis makes it clear that around $\tau=\ii\infty$ one can achieve a more pronounced hierarchy than at $\omega$ or $\ii$.
This is because for the latter two critical points the relevant part of the residual group is either $\mathds{Z}_3^{\GeneratorS\,\GeneratorT}$ or $\mathds{Z}_2^{\GeneratorS}$, which can give zeros of at most order 3 or 2, respectively. 
On the other hand, at $\ii\infty$, the relevant residual group has an order determined by the order $N$ of the $\GeneratorT$ generator representation, which can exceed 3, and thus result in a larger hierarchy.  
This point can also be directly seen from the statement~\ref{item:det_max_hierarchy} on page~\pageref{item:det_max_hierarchy}.
That is, the mass product $m_1\,m_2\,m_3$ is at most proportional to
\begin{equation}\label{eq:uEpsilonRelation}
    u^{k_{\det M}} = \varepsilon^{N\, k_{\det M} /12}\;.
\end{equation}
Notice that, according to our discussion in \Cref{sec:1-dimensional_VVMF}, the exponent of $\varepsilon$ is an integer.
For instance, in the quark sector, if $\varepsilon$ is of order $10^{-1}$, then an appropriate choice of hierarchy could be $1\relative\varepsilon^3\relative\varepsilon^5$ for the mass eigenvalues of the up quark, which requires at least a zero of order 8 for the determinant of the mass matrix. 
This suggests that for the quark sector, among all critical points, $\ii\infty$ is the only one that leads naturally to large hierarchies.

It is worth noting that, for the same finite modular group, different realizations of the modular flavor symmetries can lead to a  different order of the $\GeneratorT$ generator. 
One example is studied in \cite{Arriaga-Osante:2025ppz}, where the finite modular group $T'$ is examined, but with different presentations in which the $\GeneratorT$ generator has orders 3 or 6. 
The latter option yields a more pronounced hierarchy.

\subsection{Limitations}
\label{sec:Limitations}

A well-known, though less discussed, limitation of the scheme of modular flavor symmetries is that the \ac{EFT} is more subtle than in other schemes. 
This is evident from the pioneering paper \cite{Feruglio:2017spp}.\footnote{We thank Ferruccio Feruglio for enlightening discussions on these subtleties.} 
Models 1 and 2 of this paper lead to neutrino masses. 
While model 2 is a see-saw model, model 1 starts from the Weinberg operator. 
However, integrating out the right-handed neutrinos in model 2 does not lead to model 1. 
Rather, it leads to a model that does not comply with the rules of modular flavor model building: unlike in model 1, the low-energy Weinberg operator has singularities in the fundamental domain. 
Of course, these singularities correspond to the zeros of the right-handed neutrino mass matrix.
Close to these singularities additional hierarchies beyond those discussed in this work may arise.

This shows that an \ac{EFT} derived from a modular flavor symmetric model does, in general, not lead to a model in which all couplings are modular forms. 
Therefore, our conclusions may not apply to \acp{EFT} in which states were integrated out that may become massless in certain regions of $\tau$-modulus space.

Furthermore, as discussed in \Cref{sec:Mass_matrices_and_mass_hierarchies}, in our analysis we focus on modular weights that are not excessively large.  Another limitation is that we mainly focus on the near-critical analysis of the mass hierarchies. The near-critical analysis of the other flavor parameters, such as the mixing angles and the \CP-violating phases, is left for future work. 
In addition, the analysis in our paper is mainly based on the minimal K\"ahler potential. 
Corrections to the K\"ahler potential will not lift the zeros of the determinant of mass matrices, nor will they significantly affect our near-critical analysis. 
Of course, extreme deformations of the K\"ahler metric do have significant impact. 
On the other hand, it has been pointed out that in certain top-down models the corrections to the K\"ahler metric are parametrically small for modulus values close to the critical points \cite{Li:2025bsr}.
Therefore, there is a class of well-motivated models in which the K\"ahler corrections do not affect our conclusions.

Notice also that in \ac{UV} completions with modular flavor symmetries the critical points have additional important properties. 
Specifically, in orbifold compactifications of the heterotic string at these points extra gauge bosons become massless \cite{Dijkgraaf:1989hb,Ibanez:1990ju,Beye:2014nxa,Li:2025bsr}.
This means that for near-critical values of $\tau$ there are additional, in principle testable, predictions such as parametrically light gauge bosons~\cite{Li:2025bsr}. 
Our present analysis leads to relations between fermion mass hierarchies and the suppression of the masses of these gauge bosons.

\subsection{Modular vs.\ conventional Froggatt--Nielsen mechanisms}
\label{sec:Modular_vs_conventional_FN_mechanism}

As we have argued around \Cref{eq:generalOmegaGamma0}, the criticality expansion may be regarded as a \acf{FN} mechanism.
In the traditional \ac{FN} scheme \cite{Froggatt:1978nt}, couplings such as entries of mass matrices can be expanded in terms of
\begin{equation}
 \varepsilon_{\mathrm{FN}}\defeq \frac{\braket{\phi_\mathrm{FN}}}{\Lambda}\;, 
\end{equation}
where $\braket{\phi_\mathrm{FN}}$ denotes the \ac{VEV} of the \ac{FN} flavon and $\Lambda$ the cut-off scale. 
The leading order terms of different couplings come with different exponents of $\varepsilon_{\mathrm{FN}}$, thus explaining hierarchies. 
While the \ac{FN} mechanism is very simple and compelling, a possible drawback is that one has to make additional assumptions about the coefficients (see e.g.\ \cite{Fedele:2020fvh,Ibe:2024cvi} for a comprehensive discussion).
This introduces a large number of parameters in the models.
In scenarios in which the coefficients can be calculated, they are sometimes found to have hierarchies on their own \cite{Cvetic:1998gv}, and e.g.\ the next-to-leading term may be more important than the leading-order term.

These concerns do not apply to the modular variant since we know all the coefficients from the expansions of the modular forms. 
That is, the theory of modular forms provides us with the knowledge of all terms in the expansion. 
Even if the coefficients have not been worked out explicitly, one can find bounds on their growth (cf.\ \Cref{app:Estimate}).
Therefore, while modular flavor symmetric models are often constructed as \acp{EFT}, these \acp{EFT} are ``better'' in the sense that we have full control over the coefficients of $\tau$-dependent terms.\footnote{Note that certain limitations apply, cf.\ \Cref{sec:Limitations}.}

\subsection{Fermion mass hierarchy arising from coefficients}
\label{sec:coeff}

In the previous section, we noted that fermion mass hierarchy can naturally arise from the deviation of the modulus \ac{VEV} from some particular points. 
While the modular symmetry approach provides us with a qualitative understanding of the flavor puzzle, precise quantitative predictions also depend on coefficients in front of the expanding parameter $\varepsilon$, such as the coefficients $c_{1,2,3}$ in \Cref{eq:GeneralHierarchy}. 
In practice, these coefficients often rely on additional input parameters in the model, which are permitted by symmetries.
A concrete example is given by the charged lepton mass model in \cite{Ding:2022nzn}, where $\alpha$, $\beta$ and $\gamma$ are free parameters in addition to $\tau$. 
In these cases, large mass hierarchies may result from the fine-tuning of these free parameters. For example, in the model of \cite{Ding:2022nzn}, the best fit values $\beta\approx \sqrt{3}\,\alpha$ lead to $\det M_e\approx 0$. 
A large fraction of contemporary modular flavor models have similar features. 
While these coincidences may have an explanation, our work is focussed on explanations that relate the hierarchies to intrinsic properties of modular forms.

Our focus is on scenarios in which the coefficients, such as $c_{1,2,3}$ in \Cref{eq:GeneralHierarchy}, are fully, or at least largely, predicted by the properties of the modular forms. 
As we have seen in minimalistic sample models in \Cref{sec:Zeros_and_mass_hierarchies}, the coefficients $c_{1,2,3}$ are completely fixed without any additional input parameters. 
They are determined by the $q$-expansion coefficients of the \acp{VVMF} together with the \ac{CG} coefficients. 
This feature of modular flavor symmetries of fixing the coefficients is the biggest difference from the classic \ac{FN} scheme, and it is also the reason why the modular symmetry approach has such strong predictive power. 
For example, in the model of \Cref{subsec:Example3x3Weight3}, the mass hierarchy is determined to be $1\relative1\relative2\sqrt{2}\,\varepsilon$ with $\varepsilon=\ee^{2\pi\ii\tau/4}=u^3$ (see~\Cref{eq:HierarchyOfModel3.3}), where the coefficients $\{1,1,2\sqrt{2}\}$ are derived entirely from the leading-term coefficients in the $q$-expansion of modular form~\eqref{eq:q-expansionOfY3hat}. 
Clearly, this goes beyond the scope of near-critical analysis.
It is tempting to speculate that these coefficients will play an important role in the precise matching of experimental data. 

Mathematically, the $q$-expansion coefficients of modular forms contain profound arithmetic information. And now we have argued that these intriguing coefficients may also be relevant for the flavor observables. 
In other words, it is conceivable that the coefficients play a key role in solving the notorious flavor puzzle. 
However, this also calls for a further physical understanding of those coefficients in the future, which may point to the potential string theory structure behind modular symmetry.

\section{Conclusions}
\label{sec:conclusions}

We have systematically discussed the question of hierarchies in models with modular flavor symmetries. 
A key insight is that the determinants of fermion mass matrices are the 1-dimensional \acp{VVMF} of $\Group{SL}{2,\mathds{Z}}$. 
The latter are completely generated by polynomials of $\eta^2, E_4$ and $E_6$. 
Therefore, there are only limited options for the determinant, as shown in \Cref{tab:detModularForm}. 
In a given model, these 1-dimensional \acp{VVMF} are uniquely determined only by the determinant of the representations $\rho_{\det M}$ of the matter fields and the sum of their modular weights, $k_{\det M}$.
This provides us with a useful, model-independent analysis tool. 
As we discussed in detail in \Cref{sec:1-dimensional_VVMF}, given the mathematical properties of the 1-dimensional \acp{VVMF}, we know that for the case of $k_{\det M}<12$, the zeros of the determinant only occur at three critical points $\ii$, $\omega$ and $\ii\infty$. 
Therefore, as shown in \Cref{eq:GeneralHierarchy}, deviations of modulus \ac{VEV} from these critical points naturally generate mass hierarchies in modular flavor models. 
We find that the hierarchies near $\ii$ and $\omega$ are bounded from above at $\varepsilon^2$ and $\varepsilon^3$, respectively. 
The hierarchies near $\ii\infty$ cannot exceed the modular weight of the determinant, i.e.\ the maximal hierarchy is $u^{k_{\det M}}=\varepsilon^{N\,k_{\det M}/12}$ as in \Cref{eq:uEpsilonRelation}. 
Depending on the properties of the finite modular group, only certain hierarchies are available.  
For instance, hierarchies near $\ii\infty$ require the existence of nontrivial 1-dimensional representations, which excludes perfect groups such as $A_5$.

To further determine the hierarchical structure in \Cref{eq:GeneralHierarchy}, we need to explore the so-called near-critical behavior~\cite{Feruglio:2023mii}, which is reviewed in detail in \Cref{sec:near-critical-behavior}. 
The basic point is that the near-critical pattern of the mass matrix only depends on the stabilizer subgroup of the corresponding critical point. 
Specifically, it depends on the linear (weighted) representation $\Omega$ of the matter field under this stabilizer subgroup.
$\Omega$ determines the form of the perturbative expansion of a given mass matrix near the critical point through \Cref{eq:massMatrixTransformationRule}. 
Since the stability groups are all Abelian subgroups of $\Group{SL}{2,\mathds{Z}}$ such as $\mathds{Z}_2$ and $\mathds{Z}_3$, there are only very limited possibilities for $\Omega$. This means that the mass matrices of a large class of modular flavor models all have a common few near-critical forms, which implies the universality of near-critical behavior. 
In \Cref{sec:classification}, we systematically investigated all the near-critical patterns of the mass matrices near $\ii$ and $\omega$ under the $\rep{2}\oplus\rep{1}$ assignments, and the detailed classification results are presented in \Cref{app:classification}. 
All the possible mass hierarchies near these two critical points, $\ii$ and $\omega$, are presented in \Cref{app:masshierarchies}.

Our main conclusion is that modular flavor symmetries are a scheme the predictive power of which we yet have to fully unravel. 
Using the analytic properties of 1-dimensional \acp{VVMF}, we were able to gather a lot of important information on the model without the need of a detailed analysis. 
The overall outcome of our analysis is that, as long as the modular weights and the order of the finite modular group are not excessively large, it will be challenging to explain the observed hierarchies of the charged fermion masses entirely from properties of modular forms of $\Group{SL}{2,\mathds{Z}}$. 
There are a few immediate ways in which these conclusions may be avoided. 
One may choose a different realization as discussed in \Cref{sub:preference_for_the_fixed_point_at_ii_infty_}, go to schemes with more than one modulus, and/or combine the scheme with other mechanisms of generating hierarchies. 

We briefly discussed some limitations of our analysis in \Cref{sec:Discussion}. 
In particular, we compared the way hierarchies emerge from modular flavor symmetries with the conventional \ac{FN} mechanism.
At first glance, the near-critical behavior of the modular flavor model is rather similar to a \ac{FN} scheme. 
However, the coefficients of near-critical expansion in the modular flavor model are far better controlled as they are derived from the (Taylor or Fourier) expansion of modular forms and the \ac{CG} coefficients.
Therefore, they are often completely fixed, and furthermore related to intricate questions in, say, number theory. 
Clearly, this is significantly different from the traditional \ac{FN} and related schemes, where all the coefficients are completely free. Further explorations of these questions are left for future work.

The emergence of hierarchies from near-critical \acp{VEV} opens up intriguing avenues for future research. 
The preference for $\tau$ to reside near critical points may reflect deeper physical principles.
It is potentially linked to moduli stabilization dynamics, supersymmetry breaking, or cosmological evolution. 
It will be interesting to explore the near-critical origins of other parameters in \ac{SM}, such as the Higgs mass.

\section*{Acknowledgments}

We would like to thank Ferruccio Feruglio for useful discussions.
The work of all authors is supported in part by the National Science Foundation, under Grant No.\ PHY-2210283. 
The authors would like to express special thanks to the Mainz Institute for Theoretical Physics (MITP) of the Cluster of Excellence PRISMA$+$ (Project ID 390831469) for its hospitality and support.
Three of us (MCC, XGL \& MR) would like to thank the organizers of BCVSPIN 2024, where part of the research was performed and some results were presented.

\appendix

\section{1- and 2-dimensional irreducible representations of \texorpdfstring{$\Group{SL}{2,\mathds{Z}}$}{SL(2,Z)}}
\label{app:1and2irreps}

\begin{table}[t]
  \begin{tabular}{ccccccccccccccccccc} 
    \toprule
  $(r_1,r_2)$    & 
  $(0,\frac{1}{2})$&$(0,\frac{2}{3})$&$(\frac{1}{3},0)$&$(\frac{2}{3},\frac{1}{3})$&$(\frac{3}{4},\frac{1}{4})$&$(\frac{3}{5},\frac{2}{5})$&$(\frac{4}{5},\frac{1}{5})$&$(\frac{1}{2},\frac{1}{6})$&$(\frac{1}{6},\frac{2}{3})$
  
  \\ 
  $N$      & 2&3&3&3&4&5&5&6&6 \\ \midrule
  $(r_1,r_2)$    & $(\frac{1}{6},\frac{5}{6})$&$(\frac{5}{6},\frac{1}{2})$&$(\frac{5}{6},\frac{1}{3})$&$(\frac{1}{8},\frac{3}{8})$&$(\frac{3}{8},\frac{5}{8})$&$(\frac{5}{8},\frac{7}{8})$&$(\frac{7}{8},\frac{1}{8})$&$(\frac{1}{10},\frac{9}{10})$&$(\frac{3}{10},\frac{7}{10})$ \\ 
  $N$      & 6&6&6&8&8&8&8&10&10 \\ \midrule 
  $(r_1,r_2)$    & $(\frac{11}{12},\frac{5}{12})$&$(\frac{11}{12},\frac{7}{12})$&$(\frac{1}{12},\frac{3}{4})$&$(\frac{1}{12},\frac{7}{12})$&$(\frac{1}{4},\frac{11}{12})$&$(\frac{3}{4},\frac{5}{12})$&$(\frac{5}{12},\frac{1}{12})$&$(\frac{7}{12},\frac{1}{4})$&$(\frac{14}{15},\frac{11}{15})$ \\ 
  $N$      & 12&12&12&12&12&12&12&12&15\\ \midrule
  $(r_1,r_2)$    & $(\frac{2}{15},\frac{8}{15})$&$(\frac{4}{15},\frac{1}{15})$&$(\frac{7}{15},\frac{13}{15})$&$(\frac{11}{20},\frac{19}{20})$&$(\frac{1}{20},\frac{9}{20})$&$(\frac{17}{20},\frac{13}{20})$&$(\frac{7}{20},\frac{3}{20})$&$(\frac{11}{24},\frac{17}{24})$&$(\frac{1}{24},\frac{7}{24})$\\ 
  $N$      & 15&15&15&20&20&20&20&24&24 \\ \midrule
  $(r_1,r_2)$    & $(\frac{13}{24},\frac{19}{24})$&$(\frac{17}{24},\frac{23}{24})$&$(\frac{19}{24},\frac{1}{24})$&$(\frac{23}{24},\frac{5}{24})$&$(\frac{5}{24},\frac{11}{24})$&$(\frac{7}{24},\frac{13}{24})$&$(\frac{13}{30},\frac{7}{30})$&$(\frac{19}{30},\frac{1}{30})$&$(\frac{23}{30},\frac{17}{30})$ \\ 
  $N$      & 24&24&24&24&24&24&30&30&30 \\ \midrule
  $(r_1,r_2)$    & 
  $(\frac{29}{30},\frac{11}{30})$&$(\frac{11}{60},\frac{59}{60})$&$(\frac{13}{60},\frac{37}{60})$&$(\frac{1}{60},\frac{49}{60})$&$(\frac{23}{60},\frac{47}{60})$&$(\frac{31}{60},\frac{19}{60})$&$(\frac{41}{60},\frac{29}{60})$&$(\frac{43}{60},\frac{7}{60})$&$(\frac{53}{60},\frac{17}{60})$
  \\ 
  $N$      & 30&60&60&60&60&60&60&60&60 \\ \bottomrule
  \end{tabular}
  \caption{\label{tab:2-d_irrep}Summary of the 54 irreducible 2-dimensional representations of $\Group{SL}{2,\mathds{Z}}$ with finite image, where $N$ denotes the order of $\rho(\GeneratorT)$.}
  \end{table}

Since $\Group{SL}{2,\mathds{Z}}$ is very close to a free group, its representation theory is quite wild. 
However, in the applications to flavor physics, we typically focus on unitary irreducible representations, which often form finite groups. 
Therefore, we only focus on the unitary irreducible representations with finite image of $\Group{SL}{2,\mathds{Z}}$.
Some of the results of such representations have been given by mathematicians~\cite{tuba2001representations}, and all the results of 1- and 2-dimensional \acp{irrep} with finite image are summarized in \cite{Liu:2021gwa}. 
For the sake of convenience, we recap some of their properties in this appendix.

The modular group $\Group{SL}{2,\mathds{Z}}$ contains 12 1-dimensional \acp{irrep}, which we will denote as $\rep{1}_p$. 
They can be obtained from the relations $\GeneratorS^4=(\GeneratorS\,\GeneratorT)^3=\mathds{1}$ satisfied by the modular generators $\GeneratorS$ and $\GeneratorT$,
\begin{equation}
	\label{eq:SL2Zsinglets}
	\rep{1}_p:~~~~~	\rho_{\rep{1}_p}(\GeneratorS)=\ii^p\;,\qquad \rho_{\rep{1}_p}(\GeneratorT)=\ee^{\pi\ii p/6}\;,\qquad \rho_{\rep{1}_p}(\GeneratorS\,\GeneratorT) = \omega^p\;,
\end{equation}
where $p\in\{0,1,\dots,11\}$. 

The modular group $\Group{SL}{2,\mathds{Z}}$ contains 54 inequivalent 2-dimensional \acp{irrep} with finite image, denoted as $\rep{2}_{(r_1,r_2)}$. They can be obtained from the homomorphisms of $\Group{SL}{2,\mathds{Z}}$ to the finite subgroups of $\Group{SU}{2}$. 
The representation matrices of modular generators $\GeneratorS$ and $\GeneratorT$ can be parameterized as~\cite{mason20082}
\begin{subequations}\label{eq:2dirrepsMat}
\begin{align}
\nonumber
\rho_{\rep{2}_{(r_1,r_2)}}(\GeneratorS)&=\frac{(\lambda_1\,\lambda_2)^2}{\lambda_2-\lambda_1}\begin{pmatrix}
1 & \sqrt{-(\lambda_1\,\lambda_2)^5\,(\lambda_1-\lambda_2)^2-1} \\
\sqrt{-(\lambda_1\,\lambda_2)^5\,(\lambda_1-\lambda_2)^2-1} & -1
\end{pmatrix} \;,\\
\rho_{\rep{2}_{(r_1,r_2)}}(\GeneratorT)&=\begin{pmatrix}
\lambda_1 & 0 \\
0 & \lambda_2
\end{pmatrix}\;,
\end{align}
\end{subequations}
with $\lambda_{1,2}\defeq\ee^{2 \pi \ii r_{1,2}}$. 
The pair $(r_1,r_2)$ uniquely determines the 2-dimensional \acp{irrep} up to a similarity transformation.
Exchanging $r_1\leftrightarrow r_2$ does not affect the result. The 54 possibilities are shown in \Cref{tab:2-d_irrep}, where $N$ denotes the order of $\rho(\GeneratorT)$.

One can verify that the eigenvalues of $\rho(\GeneratorS)$ are always $\pm 1$ or $\pm \ii$, and the eigenvalues of $\rho(\GeneratorS\,\GeneratorT)$ are always two distinct elements of $\{1, \omega, \omega^2\}$. 
Therefore, in their respective diagonal bases, we always have
\begin{equation}
  \widehat{\rho}(\GeneratorS)=\ii^m\,\begin{pmatrix}
 1 & 0 \\
 0 & -1 \\
\end{pmatrix} \quad\text{and}\quad
\widehat{\rho}(\GeneratorS\,\GeneratorT)=\omega^n\,\begin{pmatrix}
1 & 0 \\
0 & \omega \\
\end{pmatrix}
\end{equation}
with $m\in\{0,1,2,3\}$ and $n\in\{0,1,2\}$.

\section{Patterns of lepton mass matrices for reducible representations}
\label{app:classification}

\begin{table}
  \centering
  \begin{tabular}{cccc}
    \toprule 
    $\begin{matrix}
      s_2 & s_1\\ 
     \end{matrix}$
     & $M_{\bar e e}$ & $M_\nu$ & $M_\nu^{-1}$ \\
    \midrule
    $
\begin{matrix}
 0 & 0 \\
 2 & 0 \\
\end{matrix}
$&$
\begin{pmatrix}
 m_{11}^{{(0)}} & \widetilde{m}_{12}^{{(1)}} \, \widetilde{\varepsilon}  & m_{13}^{{(0)}} \\
 \widetilde{m}_{21}^{{(1)}}\, \widetilde{\varepsilon}  & m_{22}^{{(0)}} &  \widetilde{m}_{23}^{{(1)}}\, \widetilde{\varepsilon} \\
 m_{31}^{{(0)}} & \widetilde{m}_{32}^{{(1)}}\, \widetilde{\varepsilon}  & m_{33}^{{(0)}} \\
\end{pmatrix}
$&$
\begin{pmatrix}
 m_{11}^{{(0)}} & m_{12}^{{(1)}}\, \varepsilon  & m_{13}^{{(0)}} \\
 m_{21}^{{(1)}}\, \varepsilon  & m_{22}^{{(0)}} & m_{23}^{{(1)}}\, \varepsilon \\
 m_{31}^{{(0)}} & m_{32}^{{(1)}}\, \varepsilon  & m_{33}^{{(0)}} \\
\end{pmatrix}
$&$
\begin{pmatrix}
 m_{11}^{{(0)}} & m_{12}^{{(1)}}\, \varepsilon  & m_{13}^{{(0)}} \\
 m_{21}^{{(1)}}\, \varepsilon  & m_{22}^{{(0)}} & m_{23}^{{(1)}}\, \varepsilon  \\
 m_{31}^{{(0)}} & m_{32}^{{(1)}}\, \varepsilon  & m_{33}^{{(0)}} \\
\end{pmatrix}
$\\ \midrule$
\begin{matrix}
 0 & 1 \\
 2 & 1 \\
\end{matrix}
$&$
\begin{pmatrix}
 m_{11}^{{(0)}} &\widetilde{m}_{12}^{{(1)}}\,  \widetilde{\varepsilon}  & 0 \\
 \widetilde{m}_{21}^{{(1)}}\, \widetilde{\varepsilon}  & m_{22}^{{(0)}} & 0 \\
 0 & 0 & m_{33}^{{(0)}} \\
\end{pmatrix}
$&$
\begin{pmatrix}
 m_{11}^{{(0)}} & m_{12}^{{(1)}}\, \varepsilon  & 0 \\
 m_{21}^{{(1)}}\, \varepsilon  & m_{22}^{{(0)}} & 0 \\
 0 & 0 & m_{33}^{{(1)}}\, \varepsilon  \\
\end{pmatrix}
$&$
\begin{pmatrix}
 m_{11}^{{(0)}} & m_{12}^{{(1)}}\, \varepsilon  & 0 \\
 m_{21}^{{(1)}}\, \varepsilon  & m_{22}^{{(0)}} & 0 \\
 0 & 0 & m_{33}^{{(1)}}\, \varepsilon  \\
\end{pmatrix}
$\\ \midrule$
\begin{matrix}
 0 & 2 \\
 2 & 2 \\
\end{matrix}
$&$
\begin{pmatrix}
 m_{11}^{{(0)}} & \widetilde{m}_{12}^{{(1)}}\, \widetilde{\varepsilon}  & \widetilde{m}_{13}^{{(1)}}\, \widetilde{\varepsilon}  \\
 \widetilde{m}_{21}^{{(1)}}\, \widetilde{\varepsilon}  & m_{22}^{{(0)}} & m_{23}^{{(0)}} \\
 \widetilde{m}_{31}^{{(1)}}\, \widetilde{\varepsilon}  & m_{32}^{{(0)}} & m_{33}^{{(0)}} \\
\end{pmatrix}
$&$
\begin{pmatrix}
 m_{11}^{{(0)}} & m_{12}^{{(1)}}\, \varepsilon  & m_{13}^{{(1)}}\, \varepsilon  \\
 m_{21}^{{(1)}}\, \varepsilon  & m_{22}^{{(0)}} & m_{23}^{{(0)}} \\
 m_{31}^{{(1)}}\, \varepsilon  & m_{32}^{{(0)}} & m_{33}^{{(0)}} \\
\end{pmatrix}
$&$
\begin{pmatrix}
 m_{11}^{{(0)}} & m_{12}^{{(1)}}\, \varepsilon  & m_{13}^{{(1)}}\, \varepsilon  \\
 m_{21}^{{(1)}}\, \varepsilon  & m_{22}^{{(0)}} & m_{23}^{{(0)}} \\
 m_{31}^{{(1)}}\, \varepsilon  & m_{32}^{{(0)}} & m_{33}^{{(0)}} \\
\end{pmatrix}
$\\ \midrule$
\begin{matrix}
 0 & 3 \\
 2 & 3 \\
\end{matrix}
$&$
\begin{pmatrix}
 m_{11}^{{(0)}} & \widetilde{m}_{12}^{{(1)}}\, \widetilde{\varepsilon}  & 0 \\
 \widetilde{m}_{21}^{{(1)}}\, \widetilde{\varepsilon}  & m_{22}^{{(0)}} & 0 \\
 0 & 0 & m_{33}^{{(0)}} \\
\end{pmatrix}
$&$
\begin{pmatrix}
 m_{11}^{{(0)}} & m_{12}^{{(1)}}\, \varepsilon  & 0 \\
 m_{21}^{{(1)}}\, \varepsilon  & m_{22}^{{(0)}} & 0 \\
 0 & 0 & m_{33}^{{(1)}}\, \varepsilon  \\
\end{pmatrix}
$&$
\begin{pmatrix}
 m_{11}^{{(0)}} & m_{12}^{{(1)}}\, \varepsilon  & 0 \\
 m_{21}^{{(1)}}\, \varepsilon  & m_{22}^{{(0)}} & 0 \\
 0 & 0 & m_{33}^{{(1)}}\, \varepsilon  \\
\end{pmatrix}
$\\ \midrule$
\begin{matrix}
 1 & 0 \\
 3 & 0 \\
\end{matrix}
$&$
\begin{pmatrix}
 m_{11}^{{(0)}} & \widetilde{m}_{12}^{{(1)}}\, \widetilde{\varepsilon}  & 0 \\
 \widetilde{m}_{21}^{{(1)}}\, \widetilde{\varepsilon}  & m_{22}^{{(0)}} & 0 \\
 0 & 0 & m_{33}^{{(0)}} \\
\end{pmatrix}
$&$
\begin{pmatrix}
 m_{11}^{{(1)}}\, \varepsilon  & m_{12}^{{(0)}} & 0 \\
 m_{21}^{{(0)}} & m_{22}^{{(1)}}\, \varepsilon  & 0 \\
 0 & 0 & m_{33}^{{(0)}} \\
\end{pmatrix}
$&$
\begin{pmatrix}
 m_{11}^{{(1)}}\, \varepsilon  & m_{12}^{{(0)}} & 0 \\
 m_{21}^{{(0)}} & m_{22}^{{(1)}}\, \varepsilon  & 0 \\
 0 & 0 & m_{33}^{{(0)}} \\
\end{pmatrix}
$\\ \midrule$
\begin{matrix}
 1 & 1 \\
 3 & 1 \\
\end{matrix}
$&$
\begin{pmatrix}
 m_{11}^{{(0)}} & \widetilde{m}_{12}^{{(1)}}\, \widetilde{\varepsilon}  & m_{13}^{{(0)}} \\
 \widetilde{m}_{21}^{{(1)}}\, \widetilde{\varepsilon}  & m_{22}^{{(0)}} & \widetilde{m}_{23}^{{(1)}}\, \widetilde{\varepsilon}  \\
 m_{31}^{{(0)}} & \widetilde{m}_{32}^{{(1)}}\, \widetilde{\varepsilon}  & m_{33}^{{(0)}} \\
\end{pmatrix}
$&$
\begin{pmatrix}
 m_{11}^{{(1)}}\, \varepsilon  & m_{12}^{{(0)}} & m_{13}^{{(1)}}\, \varepsilon  \\
 m_{21}^{{(0)}} & m_{22}^{{(1)}}\, \varepsilon  & m_{23}^{{(0)}} \\
 m_{31}^{{(1)}}\, \varepsilon  & m_{32}^{{(0)}} & m_{33}^{{(1)}}\, \varepsilon  \\
\end{pmatrix}
$&$\Boxed{
\begin{pmatrix}
 m_{11}^{{(1)}}\, \varepsilon  & m_{12}^{{(0)}} & m_{13}^{{(1)}}\, \varepsilon  \\
 m_{21}^{{(0)}} & m_{22}^{{(1)}}\, \varepsilon  & m_{23}^{{(0)}} \\
 m_{31}^{{(1)}}\, \varepsilon  & m_{32}^{{(0)}} & m_{33}^{{(1)}}\, \varepsilon  \\
\end{pmatrix}}
$\\ \midrule$
\begin{matrix}
 1 & 2 \\
 3 & 2 \\
\end{matrix}
$&$
\begin{pmatrix}
 m_{11}^{{(0)}} & \widetilde{m}_{12}^{{(1)}}\, \widetilde{\varepsilon}  & 0 \\
 \widetilde{m}_{21}^{{(1)}}\, \widetilde{\varepsilon}  & m_{22}^{{(0)}} & 0 \\
 0 & 0 & m_{33}^{{(0)}} \\
\end{pmatrix}
$&$
\begin{pmatrix}
 m_{11}^{{(1)}}\, \varepsilon  & m_{12}^{{(0)}} & 0 \\
 m_{21}^{{(0)}} & m_{22}^{{(1)}}\, \varepsilon  & 0 \\
 0 & 0 & m_{33}^{{(0)}} \\
\end{pmatrix}
$&$
\begin{pmatrix}
 m_{11}^{{(1)}}\, \varepsilon  & m_{12}^{{(0)}} & 0 \\
 m_{21}^{{(0)}} & m_{22}^{{(1)}}\, \varepsilon  & 0 \\
 0 & 0 & m_{33}^{{(0)}} \\
\end{pmatrix}
$\\ \midrule$
\begin{matrix}
 1 & 3 \\
 3 & 3 \\
\end{matrix}
$&$
\begin{pmatrix}
 m_{11}^{{(0)}} & \widetilde{m}_{12}^{{(1)}}\, \widetilde{\varepsilon}  & \widetilde{m}_{13}^{{(1)}}\, \widetilde{\varepsilon}  \\
 \widetilde{m}_{21}^{{(1)}}\, \widetilde{\varepsilon}  & m_{22}^{{(0)}} & m_{23}^{{(0)}} \\
 \widetilde{m}_{31}^{{(1)}}\, \widetilde{\varepsilon}  & m_{32}^{{(0)}} & m_{33}^{{(0)}} \\
\end{pmatrix}
$&$
\begin{pmatrix}
 m_{11}^{{(1)}}\, \varepsilon  & m_{12}^{{(0)}} & m_{13}^{{(0)}} \\
 m_{21}^{{(0)}} & m_{22}^{{(1)}}\, \varepsilon  & m_{23}^{{(1)}}\, \varepsilon  \\
 m_{31}^{{(0)}} & m_{32}^{{(1)}}\, \varepsilon  & m_{33}^{{(1)}}\, \varepsilon  \\
\end{pmatrix}
$&$\Boxed{
\begin{pmatrix}
 m_{11}^{{(1)}}\, \varepsilon  & m_{12}^{{(0)}} & m_{13}^{{(0)}} \\
 m_{21}^{{(0)}} & m_{22}^{{(1)}}\, \varepsilon  & m_{23}^{{(1)}}\, \varepsilon  \\
 m_{31}^{{(0)}} & m_{32}^{{(1)}}\, \varepsilon  & m_{33}^{{(1)}}\, \varepsilon  \\
\end{pmatrix}}
$\\
    \bottomrule
   \end{tabular}
  \caption{\label{tab:invS2+1MassMatrix}\TpO\ representation mass matrix pattern near the fixed point $\tau_0 = \ii$ with a subgroup generated by $\GeneratorS$ in the leading order. 
  We have defined $\widetilde{m}^{(1)}_{ij}\, \widetilde{\varepsilon} \defeq  m^{(10)}_{ij}\, \varepsilon +  m^{(01)}_{ij}\, \bar \varepsilon$. 
  Notice that \Cref{eq:generalOmega2+1} has some redundancy, as the actual degrees of freedom are the overall phase and the relative phase between the two irreducible representations. 
  Therefore, for $s_2 = 2$ or $3$, the result would be the same as $s_2 = 0$ or $1$ if we permute the first two diagonal elements in $\Omega$.
  Adding a nontrivial $\Omega_{\mathscr{W}}$ as in \Cref{eq:massMatrixTransformationRule} does not result in additional matrix patterns. 
  The boxes highlight $M_\nu^{-1}$ patterns which agree with experimental data without fine-tuning.} 
\end{table}
\begin{table}
  \centering
  \begin{tabular}{cccc}
    \toprule
    $\begin{matrix}
      s_2 & s_1\\ 
     \end{matrix}$
      & $M_{\bar e e}$ & $M_\nu$ & $M_\nu^{-1}$ \\
    \midrule
     $
\begin{matrix}
 0 & 0 \\
\end{matrix}
$&$
\begin{pmatrix}
 m_{11}^{{(0)}} & m_{12}^{{(1)}}\, \varepsilon  & m_{13}^{{(0)}} \\
 m_{21}^{{(01)}}\, \bar{\varepsilon}  & m_{22}^{{(0)}} & m_{23}^{{(01)}}\, \bar{\varepsilon}  \\
 m_{31}^{{(0)}} & m_{32}^{{(1)}}\, \varepsilon  & m_{33}^{{(0)}} \\
\end{pmatrix}
$&$
\begin{pmatrix}
 m_{11}^{{(0)}} & m_{12}^{{(1)}}\, \varepsilon  & m_{13}^{{(0)}} \\
 m_{21}^{{(1)}}\, \varepsilon  & m_{22}^{{(2)}}\, \varepsilon^2  & m_{23}^{{(1)}}\, \varepsilon  \\
 m_{31}^{{(0)}} & m_{32}^{{(1)}}\, \varepsilon  & m_{33}^{{(0)}} \\
\end{pmatrix}
$&$
\begin{pmatrix}
 m_{11}^{{(0)}} & m_{12}^{{(2)}}\, \varepsilon^2  & m_{13}^{{(0)}} \\
 m_{21}^{{(2)}}\, \varepsilon^2 & m_{22}^{{(1)}}\, \varepsilon  & m_{23}^{{(2)}}\, \varepsilon^2  \\
 m_{31}^{{(0)}} & m_{32}^{{(2)}}\, \varepsilon^2  & m_{33}^{{(0)}} \\
\end{pmatrix}
$\\ \midrule$
\begin{matrix}
 0 & 1 \\
\end{matrix}
$&$
\begin{pmatrix}
 m_{11}^{{(0)}} & m_{12}^{{(1)}}\, \varepsilon  & m_{13}^{{(1)}}\, \varepsilon  \\
 m_{21}^{{(01)}}\, \bar{\varepsilon}  & m_{22}^{{(0)}} & m_{23}^{{(0)}} \\
 m_{31}^{{(01)}}\, \bar{\varepsilon}  & m_{32}^{{(0)}} & m_{33}^{{(0)}} \\
\end{pmatrix}
$&$
\begin{pmatrix}
 m_{11}^{{(0)}} & m_{12}^{{(1)}}\, \varepsilon  & m_{13}^{{(1)}}\, \varepsilon  \\
 m_{21}^{{(1)}}\, \varepsilon  & m_{22}^{{(2)}}\, \varepsilon^2  & m_{23}^{{(2)}}\, \varepsilon^2  \\
 m_{31}^{{(1)}}\, \varepsilon  & m_{32}^{{(2)}}\, \varepsilon^2  & m_{33}^{{(2)}}\, \varepsilon^2  \\
\end{pmatrix}
$&$
\begin{pmatrix}
 m_{11}^{{(0)}} & m_{12}^{{(2)}}\, \varepsilon^2  & m_{13}^{{(2)}}\, \varepsilon^2  \\
 m_{21}^{{(2)}}\, \varepsilon^2  & m_{22}^{{(1)}}\, \varepsilon  & m_{23}^{{(1)}}\, \varepsilon  \\
 m_{31}^{{(2)}}\, \varepsilon^2  & m_{32}^{{(1)}}\, \varepsilon  & m_{33}^{{(1)}}\, \varepsilon  \\
\end{pmatrix}
$\\ \midrule$
\begin{matrix}
 0 & 2 \\
 1 & 0 \\
 2 & 1 \\
\end{matrix}
$&$
\begin{pmatrix}
 m_{11}^{{(0)}} & m_{12}^{{(1)}}\, \varepsilon  & m_{13}^{{(01)}}\, \bar{\varepsilon}  \\
 m_{21}^{{(01)}}\, \bar{\varepsilon}  & m_{22}^{{(0)}} & m_{23}^{{(1)}}\, \varepsilon  \\
 m_{31}^{{(1)}}\, \varepsilon  & m_{32}^{{(01)}}\, \bar{\varepsilon}  & m_{33}^{{(0)}} \\
\end{pmatrix}
$&$
\begin{pmatrix}
 m_{11}^{{(0)}} & m_{12}^{{(1)}}\, \varepsilon  & m_{13}^{{(2)}}\, \varepsilon^2  \\
 m_{21}^{{(1)}}\, \varepsilon  & m_{22}^{{(2)}}\, \varepsilon^2  & m_{23}^{{(0)}} \\
 m_{31}^{{(2)}}\, \varepsilon^2  & m_{32}^{{(0)}} & m_{33}^{{(1)}}\, \varepsilon  \\
\end{pmatrix}
$&$
\begin{pmatrix}
 m_{11}^{{(0)}} & m_{12}^{{(2)}}\, \varepsilon^2  & m_{13}^{{(1)}}\, \varepsilon  \\
 m_{21}^{{(2)}}\, \varepsilon^2  & m_{22}^{{(1)}}\, \varepsilon  & m_{23}^{{(0)}} \\
 m_{31}^{{(1)}}\, \varepsilon  & m_{32}^{{(0)}} & m_{33}^{{(2)}}\, \varepsilon^2  \\
\end{pmatrix}
$\\ \midrule$
\begin{matrix}
 1 & 1 \\
\end{matrix}
$&$
\begin{pmatrix}
 m_{11}^{{(0)}} & m_{12}^{{(1)}}\, \varepsilon  & m_{13}^{{(0)}} \\
 m_{21}^{{(01)}}\, \bar{\varepsilon}  & m_{22}^{{(0)}} & m_{23}^{{(01)}}\, \bar{\varepsilon}  \\
 m_{31}^{{(0)}} & m_{32}^{{(1)}}\, \varepsilon  & m_{33}^{{(0)}} \\
\end{pmatrix}
$&$
\begin{pmatrix}
 m_{11}^{{(2)}}\, \varepsilon^2  & m_{12}^{{(0)}} & m_{13}^{{(2)}}\, \varepsilon^2  \\
 m_{21}^{{(0)}} & m_{22}^{{(1)}}\, \varepsilon  & m_{23}^{{(0)}} \\
 m_{31}^{{(2)}}\, \varepsilon^2  & m_{32}^{{(0)}} & m_{33}^{{(2)}}\, \varepsilon^2  \\
\end{pmatrix}
$&$
\Boxed{
\begin{pmatrix}
 m_{11}^{{(1)}}\, \varepsilon  & m_{12}^{{(0)}} & m_{13}^{{(1)}}\, \varepsilon  \\
 m_{21}^{{(0)}} & m_{22}^{{(2)}}\, \varepsilon^2  & m_{23}^{{(0)}} \\
 m_{31}^{{(1)}}\, \varepsilon  & m_{32}^{{(0)}} & m_{33}^{{(1)}}\, \varepsilon  \\
\end{pmatrix}}
$\\ \midrule$
\begin{matrix}
 1 & 2 \\
\end{matrix}
$&$
\begin{pmatrix}
 m_{11}^{{(0)}} & m_{12}^{{(1)}}\, \varepsilon  & m_{13}^{{(1)}}\, \varepsilon  \\
 m_{21}^{{(01)}}\, \bar{\varepsilon}  & m_{22}^{{(0)}} & m_{23}^{{(0)}} \\
 m_{31}^{{(01)}}\, \bar{\varepsilon}  & m_{32}^{{(0)}} & m_{33}^{{(0)}} \\
\end{pmatrix}
$&$
\begin{pmatrix}
 m_{11}^{{(2)}}\, \varepsilon^2  & m_{12}^{{(0)}} & m_{13}^{{(0)}} \\
 m_{21}^{{(0)}} & m_{22}^{{(1)}}\, \varepsilon  & m_{23}^{{(1)}}\, \varepsilon  \\
 m_{31}^{{(0)}} & m_{32}^{{(1)}}\, \varepsilon  & m_{33}^{{(1)}}\, \varepsilon  \\
\end{pmatrix}
$&$
\Boxed{
\begin{pmatrix}
 m_{11}^{{(1)}}\, \varepsilon  & m_{12}^{{(0)}} & m_{13}^{{(0)}} \\
 m_{21}^{{(0)}} & m_{22}^{{(2)}}\, \varepsilon^2  & m_{23}^{{(2)}}\, \varepsilon^2  \\
 m_{31}^{{(0)}} & m_{32}^{{(2)}}\, \varepsilon^2  & m_{33}^{{(2)}}\, \varepsilon^2  \\
\end{pmatrix}}
$\\ \midrule$
\begin{matrix}
 2 & 0 \\
\end{matrix}
$&$
\begin{pmatrix}
 m_{11}^{{(0)}} & m_{12}^{{(1)}}\, \varepsilon  & m_{13}^{{(1)}}\, \varepsilon  \\
 m_{21}^{{(01)}}\, \bar{\varepsilon}  & m_{22}^{{(0)}} & m_{23}^{{(0)}} \\
 m_{31}^{{(01)}}\, \bar{\varepsilon}  & m_{32}^{{(0)}} & m_{33}^{{(0)}} \\
\end{pmatrix}
$&$
\begin{pmatrix}
 m_{11}^{{(1)}}\, \varepsilon  & m_{12}^{{(2)}}\, \varepsilon^2  & m_{13}^{{(2)}}\, \varepsilon^2  \\
 m_{21}^{{(2)}}\, \varepsilon^2  & m_{22}^{{(0)}} & m_{23}^{{(0)}} \\
 m_{31}^{{(2)}}\, \varepsilon^2  & m_{32}^{{(0)}} & m_{33}^{{(0)}} \\
\end{pmatrix}
$&$
\begin{pmatrix}
 m_{11}^{{(2)}}\, \varepsilon^2  & m_{12}^{{(1)}}\, \varepsilon  & m_{13}^{{(1)}}\, \varepsilon  \\
 m_{21}^{{(1)}}\, \varepsilon  & m_{22}^{{(0)}} & m_{23}^{{(0)}} \\
 m_{31}^{{(1)}}\, \varepsilon  & m_{32}^{{(0)}} & m_{33}^{{(0)}} \\
\end{pmatrix}
$\\ \midrule$
\begin{matrix}
 2 & 2 \\
\end{matrix}
$&$
\begin{pmatrix}
 m_{11}^{{(0)}} & m_{12}^{{(1)}}\, \varepsilon  & m_{13}^{{(0)}} \\
 m_{21}^{{(01)}}\, \bar{\varepsilon}  & m_{22}^{{(0)}} & m_{23}^{{(01)}}\, \bar{\varepsilon}  \\
 m_{31}^{{(0)}} & m_{32}^{{(1)}}\, \varepsilon  & m_{33}^{{(0)}} \\
\end{pmatrix}
$&$
\begin{pmatrix}
 m_{11}^{{(1)}}\, \varepsilon  & m_{12}^{{(2)}}\, \varepsilon^2  & m_{13}^{{(1)}}\, \varepsilon  \\
 m_{21}^{{(2)}}\, \varepsilon^2  & m_{22}^{{(0)}} & m_{23}^{{(2)}}\, \varepsilon^2  \\
 m_{31}^{{(1)}}\, \varepsilon  & m_{32}^{{(2)}}\, \varepsilon^2  & m_{33}^{{(1)}}\, \varepsilon  \\
\end{pmatrix}
$&$
\begin{pmatrix}
 m_{11}^{{(2)}}\, \varepsilon^2  & m_{12}^{{(1)}}\, \varepsilon  & m_{13}^{{(2)}}\, \varepsilon^2  \\
 m_{21}^{{(1)}}\, \varepsilon  & m_{22}^{{(0)}} & m_{23}^{{(1)}}\, \varepsilon  \\
 m_{31}^{{(2)}}\, \varepsilon^2  & m_{32}^{{(1)}}\, \varepsilon  & m_{33}^{{(2)}}\, \varepsilon^2  \\
\end{pmatrix}
$\\

  \bottomrule
  \end{tabular}
  \caption{\label{tab:invST2+1MassMatrix}\TpO\ representation mass matrix pattern near the fixed point $\tau_0 = \omega$ with a subgroup generated by $\GeneratorS\,\GeneratorT$ in the leading order. 
  The other generator $\GeneratorS^2$ excludes certain terms from the Lagrange density, and results in blocks of texture zeros in addition to the mass matrix pattern presented here. 
  Adding a nontrivial $\Omega_{\mathscr{W}}$ as in \Cref{eq:massMatrixTransformationRule} does not result in additional matrix patterns. 
  The case $(s_2,s_1) = (0,2),\ (1,0),\ (2,1)$ gives the same weighted representation of the irreducible representation and adding a nontrivial $\Omega_{\mathscr{W}}$ gives back the same mass matrix pattern up to a permutation. 
  In order to recover the correct mixing angle, i.e.\ a smaller $\sin\theta_{13}^2$ than the other two mixing angles, we require at least two order-1 off-diagonal terms. 
  The matrix having this feature is marked by a box.}
  \end{table}

In this appendix, we provide a comprehensive classification of patterns of the leptonic mass matrices in the modular flavor models based on their near-critical behavior. 
As we explained in the main text, they are determined merely by the weighted representation $\Omega(\gamma_0)$ in \Cref{eq:generalOmegaGamma0}. 
Therefore, the resulting classification essentially depend on the classification of these $\Omega(\gamma_0)$ under various assignments.

In \cite{Feruglio:2023mii}, the results for 3-dimensional irreducible representations of $\GeneratorS$, corresponding to the fixed point $\ii$, and $\GeneratorS\,\GeneratorT$, corresponding to the fixed point $\omega$, were presented. 
In these cases, $\Omega$ is completely fixed up to an overall phase factor,
\begin{subequations}
\begin{align}
  \Omega(\GeneratorS)&=\ii ^{s}\diag (1,-1,-1)\;,\\ 
  \Omega(\GeneratorS\,\GeneratorT)&=\diag(1,\omega,\omega^2) \;,
\end{align}  
\end{subequations}
where $s\in \{0,1,2,3\}$ depends on $k_{H_u}$, $k_L$ and the representation $\rho_L$.

We can construct any weighted representation in a reducible representation as the direct sum of irreducible representations, where irreducible representations of arbitrary dimensions can be chosen and fixed, up to an overall phase, in a similar manner. For the $\TpO$ representation assignment, the weighted representation matrices are, up to a permutation of the diagonal elements, given by
\begin{subequations}\label{eq:generalOmega2+1}
\begin{align}
\Omega(\GeneratorS) &= \ii^{s_{2}}\,\diag (1,-1,0) + \ii^{s_{1}}\,\diag (0,0,1) \;,\label{eq:52a} \\
    \Omega(\GeneratorS\,\GeneratorT) &= \omega^{s_{2}}\,\diag (1,\omega,0) + \omega^{s_{1}}\,\diag (0,0,1)\;,\label{eq:52b}\\
    \Omega(\GeneratorT) &= \diag \bigl(\ee^{2 \pi \ii r_1 },\ee^{2 \pi \ii r_2},0\bigr) +\diag \bigl(0,0, \ee^{ \pi \ii p/6}\bigr)\;,\\
    \Omega\bigl(\GeneratorS^2\bigr) &= (-1)^{s_{2}}\,\diag (1,1,0) + (-1)^{s_{1}}\,\diag (0,0,1) \;.\label{eq:52d}
\end{align}
\end{subequations}
Here, the labels $s_i$ take values $s_i\in\{0,1,2,3\}$ in \eqref{eq:52a}, $s_i\in\{0,1,2\}$  in \eqref{eq:52b}, and $s_i\in\{0,1\}$ in \eqref{eq:52d}, respectively.
They are determined by the modular weights $k_{H_{u,i}}, k_{L_i}$ and the representation $\rho_L$ through \Cref{eq:36b}. 
The integers $p$ take values between 0 and 11. 
We use $s_2$ for the 2-dimensional irreducible representation and $s_1$ for the 1-dimensional representation. 
These representation decompositions therefore classify the mass matrix patterns under each critical point into different forms.

For the fixed point $\ii \infty$, the overall phase is of the form $1^s = 1$, and we can use the results for the irreducible representations of dimensions 1 and 2 (see \Cref{app:1and2irreps}), where $r_1$ and $r_2$ are chosen appropriately, as shown in \Cref{tab:2-d_irrep}.   
Now note that the stabilizer group depends on the level $N$ of the normal subgroup we are working with, which determines the order of the stabilizer $\mathds{Z}_N$. 
This implies that there are many possibilities around the fixed point $\ii \infty$.  
For the $\TpO$ scenario, the total number of possible $\GeneratorT$ generators, i.e.\ the values of the triple data $(r_1,r_2,p)$, is $12\times 54=648$ according to \Cref{app:1and2irreps}. 
Therefore, in this work, we focus only on the points $\ii$ and $\omega$, as they lead to a manageable number of classes of mass matrix patterns independent of the specific model.

If we are working in an $\rep{1}\oplus\rep{1}\oplus\rep{1}$ scenario, we can pick each element on the diagonal to be the corresponding phase as \Cref{eq:generalOmegaGamma0}. This leads to largely unconstrained scenarios the analysis of which is beyond the scope of this work.
We therefore focus on the mass matrix patterns near the fixed points $\ii$ and $\omega$ for the $\TpO$ representation.

\subsection{Mass matrix patterns for neutrinos}
\label{eq:NeutrinoMassMatrix2+1}

By using \Cref{eq:massMatrixTransformationRule,eq:generalOmega2+1} and expanding $M$ according to $\varepsilon$ as shown in \Cref{eq:Expansion}, and then taking the limit $\varepsilon \rightarrow 0$, we obtain the universal critical behavior for a given mass matrix. 
We present all possible neutrino mass matrix patterns near the critical points $\ii$ and $\omega$ in \Cref{tab:invS2+1MassMatrix} and \Cref{tab:invST2+1MassMatrix}, respectively. 
Note that the generator $\GeneratorS^2$ gives the stabilizer group $\mathds{Z}_2^{\GeneratorS^2}$ at  generic point across all of $\mathcal{F}$.  
Therefore, it is common to all stabilizer groups. 
This stabilizer group acts like parity, excluding certain terms from the Lagrange density and leads to blocks of texture zeros in addition to the presented mass matrix patterns. 
In other words, some of the coefficients $m^{(n)}_{ij}$ in those near-critical patterns in \Cref{tab:invS2+1MassMatrix} and \Cref{tab:invST2+1MassMatrix} might be constantly zero due to the effect of $\mathds{Z}_2^{S^2}$. 

In the case of local \ac{SUSY}, as mentioned in \Cref{sec:critical}, there is a nontrivial $\Omega_{\mathscr{W}}$ in \Cref{eq:massMatrixTransformationRule}. 
This additional phase shifts the power of $\varepsilon$; for example, a leading term $m^{(0)}$ is shifted to $m^{(1)}\,\varepsilon$ in the mass matrix pattern near $\ii$. 
However, this does not result in any new mass matrix patterns but rather a permutation of the existing patterns with different values of $s$. 
Moreover, in the case of irreducible representations, it results in a mass matrix pattern derived from the irreducible representation with a different $s$.

Observing \Cref{tab:invS2+1MassMatrix}, where we consider the region around the fixed point $\tau_0 = \ii$ with residual symmetry $G_0$ generated by $\GeneratorS$, all the matrix patterns can be permuted into a similar form as those in reference~\cite{Feruglio:2023mii} or exhibit trivial mixing angles.  
This means that there are no new mass matrix patterns near the fixed point $\ii$ that match the experimental data. The case of $(s_2,s_1) = (1,3)$ was identified by Feruglio in the irreducible triplet $L$ case~\cite{Feruglio:2023bav}, which is consistent with the experimental data. Another case, $(s_2,s_1) = (1,1)$, is simply a permutation of that.

However, from \Cref{tab:invST2+1MassMatrix}, we observe that there are several new mass matrix patterns near the fixed point $\omega$ that do not appear when $L$ is in the 3-dimensional irreducible representation. 
Nevertheless, most of these patterns are unlikely to produce the desired mixing angles. Experimental data indicates that we expect two relatively large mixing angles for $\theta_{12}$ and $\theta_{23}$, and one small mixing angle for $\theta_{13}$, meaning we need two off-diagonal entries of order 1.  
Only the cases $(s_2,s_1) = (1,2)$ and $(s_2,s_1) = (1,1)$ 
exhibit this feature, showing a similar pattern to the $\tau_0 = \ii$ case with $(s_2,s_1) = (1,3)$.

The eigenvalues of the mass matrix $M_\nu$ give rise to a hierarchy of neutrino masses in the ratio $1\relative 1\relative \frac{1}{\varepsilon}$, where $\varepsilon$ is small. 
This is loosely consistent with a normal hierarchy. 
One can see that this case is almost identical to the $(s_2,s_1) = (1,3)$ case in \Cref{tab:invS2+1MassMatrix}, except that some of the order-1 coefficients are zero. 
It turns out that the same diagonalization procedure as in reference~\cite{Feruglio:2023mii} can be applied to obtain similar flavor observables, producing neutrino masses and mixing angles consistent with experimental data. 

\subsection{Mass hierarchy for Dirac fermions}
\label{app:masshierarchies}

  \begin{table}[th!]
  \centering
  \begin{tabular}[t]{@{}ccc@{}} 
    \toprule
    $\Omega$  & $\Omega_\ChargeC$ & Hierarchy \\ 
    \midrule
    $\diag (-1, -1, 1)$ & $\diag (-1, -1, 1)$ & $(1,1,1)$ \\
    $\diag (-1, -\ii, \ii)$ & $\diag (-1, -\ii, \ii)$ & $(1,1,1)$ \\
    $\diag (-1, -\ii, 1)$ & $\diag (-1, \ii, 1)$ & $(1,1,1)$ \\
    $\diag (-1, 1, 1)$ & $\diag (-1, 1, 1)$ & $(1,1,1)$ \\
    $\diag (-\ii, -\ii, \ii)$ & $\diag (-\ii, \ii, \ii)$ & $(1,1,1)$ \\
    $\diag (-\ii, \ii, 1)$ & $\diag (-\ii, \ii, 1)$ & $(1,1,1)$ \\
    \midrule
    $\diag (-1, -1, 1)$ & $\diag (-1, 1, 1)$ & $(\varepsilon,1,1)$ \\
    $\diag (-1, -\ii, \ii)$ & $\diag (-\ii, \ii, 1)$ & $(\varepsilon,1,1)$ \\
    $\diag (-1, -\ii, 1)$ & $\diag (-1, -\ii, 1)$ & $(\varepsilon,1,1)$ \\
    $\diag (-1, \ii, 1)$ & $\diag (-1, \ii, 1)$ & $(\varepsilon,1,1)$ \\
    $\diag (-\ii, -\ii, \ii)$ & $\diag (-\ii, -\ii, \ii)$ & $(\varepsilon,1,1)$ \\
    $\diag (-\ii, \ii, \ii)$ & $\diag (-\ii, \ii, \ii)$ & $(\varepsilon,1,1)$ \\
    \bottomrule
  \end{tabular}
  \caption{\label{tab:chargedLeptonHierarchyI}All possible mass hierarchies for the case of Dirac fermions near the critical point $\ii$ for the $\TpO$ assignment. 
  $\varepsilon\defeq \lvert \frac{\tau-\ii}{\tau+\ii} \rvert$ measures the deviation from $\ii$.}
  \end{table}

\begin{table}[th!]
  \centering
  \begin{tabular}[t]{@{}ccc@{}} 
    \toprule
    $\Omega$  & $\Omega_\ChargeC$ & Hierarchy \\ 
    \midrule
    $\diag (1,\omega,1)$ & $\diag (1,\omega,\omega^2)$ & $(\varepsilon,1,1)$ \\
    $\diag (1,\omega,\omega)$ & $\diag (\omega,\omega^2,\omega^2)$ & $(\varepsilon,1,1)$ \\
    $\diag (1,\omega,\omega)$ & $\diag (\omega^2,1,1)$ & $(\varepsilon,1,1)$ \\
    $\diag (1,\omega,\omega^2)$ & $\diag (\omega,\omega^2,\omega)$ & $(\varepsilon,1,1)$ \\
    $\diag (1,\omega,\omega^2)$ & $\diag (\omega^2,1,\omega^2)$ & $(\varepsilon,1,1)$ \\
    $\diag (\omega,\omega^2,\omega^2)$ & $\diag (\omega,\omega^2,\omega^2)$ & $(\varepsilon,1,1)$ \\
    $\diag (\omega^2,1,1)$ & $\diag (\omega^2,1,1)$ & $(\varepsilon,1,1)$ \\\midrule
    $\diag (1,\omega,1)$ & $\diag (1,\omega,1)$ & $(\varepsilon ^2,1,1)$ \\
    $\diag (1,\omega,1)$ & $\diag (\omega^2,1,\omega^2)$ & $(\varepsilon ^2,1,1)$ \\
    $\diag (1,\omega,\omega)$ & $\diag (1,\omega,\omega^2)$ & $(\varepsilon ^2,1,1)$ \\
    $\diag (1,\omega,\omega^2)$ & $\diag (\omega,\omega^2,\omega^2)$ & $(\varepsilon ^2,1,1)$ \\
    $\diag (1,\omega,\omega^2)$ & $\diag (\omega^2,1,1)$ & $(\varepsilon ^2,1,1)$ \\
    $\diag (\omega,\omega^2,\omega)$ & $\diag (\omega,\omega^2,\omega)$ & $(\varepsilon ^2,1,1)$ \\
    $\diag (\omega,\omega^2,\omega)$ & $\diag (\omega^2,1,\omega^2)$ & $(\varepsilon ^2,1,1)$ \\\bottomrule
  \end{tabular}
  \hfill
  \begin{tabular}[t]{@{}ccc@{}} 
    \toprule
    $\Omega$  & $\Omega_\ChargeC$ & Hierarchy \\ 
    \midrule
     $\diag (1,\omega,1)$ & $\diag (\omega^2,1,1)$ & $(1,1,1)$ \\
    $\diag (1,\omega,\omega)$ & $\diag (\omega^2,1,\omega^2)$ & $(1,1,1)$ \\
    $\diag (1,\omega,\omega^2)$ & $\diag (1,\omega,\omega^2)$ & $(1,1,1)$ \\
    $\diag (\omega,\omega^2,\omega)$ & $\diag (\omega,\omega^2,\omega^2)$ & $(1,1,1)$ \\\midrule
    $\diag (1,\omega,1)$ & $\diag (\omega,\omega^2,\omega)$ & $(\varepsilon ,\varepsilon ,1)$ \\
    $\diag (\omega^2,1,\omega^2)$ & $\diag (\omega^2,1,\omega^2)$ & $(\varepsilon ,\varepsilon ,1)$ \\\midrule
    $\diag (1,\omega,1)$ & $\diag (1,\omega,\omega)$ & $(\varepsilon^2 ,\varepsilon,1)$ \\
    $\diag (1,\omega,1)$ & $\diag (\omega,\omega^2,\omega^2)$ & $(\varepsilon^2 ,\varepsilon,1)$ \\
    $\diag (1,\omega,\omega)$ & $\diag (\omega,\omega^2,\omega)$ & $(\varepsilon^2 ,\varepsilon,1)$ \\
    $\diag (\omega,\omega^2,\omega)$ & $\diag (\omega^2,1,1)$ & $(\varepsilon^2 ,\varepsilon,1)$ \\
    $\diag (\omega,\omega^2,\omega^2)$ & $\diag (\omega^2,1,\omega^2)$ & $(\varepsilon^2 ,\varepsilon,1)$ \\
    $\diag (\omega^2,1,1)$ & $\diag (\omega^2,1,\omega^2)$ & $(\varepsilon^2 ,\varepsilon,1)$ \\\midrule
    $\diag (1,\omega,\omega)$ & $\diag (1,\omega,\omega)$ & $(\varepsilon ^2,\varepsilon ^2,1)$ \\
    $\diag (\omega,\omega^2,\omega^2)$ & $\diag (\omega^2,1,1)$ & $(\varepsilon ^2,\varepsilon ^2,1)$ \\
    \bottomrule
  \end{tabular}
  \caption{\label{tab:chargedLeptonHierarchy}All possible Dirac fermion hierarchies near the critical point $\omega$ for the $\TpO$ assignment. 
  $\varepsilon\defeq \lvert \frac{\tau-\omega}{\tau-\omega^2} \rvert$ measures the deviation from $\omega$.}
  \end{table}

To determine the mass matrix for Dirac fermions, such as charged leptons, the assignment of the right-handed field $E^\ChargeC$ is also needed. 
Moreover, since $M_e$ is not a Hermitean matrix, different permutations of the diagonal elements can lead to different patterns.

As before, we focus on those generators distinct from $\GeneratorS^2$, resulting in
only a $\mathds{Z}_2$ stabilizer group around $\ii$. 
This implies that the mass matrix contains only the zero-order and first-order terms in $\varepsilon$, suggesting that the smallness of $\varepsilon$ alone cannot produce the desired hierarchy for the charged lepton sector. In contrast, the fixed point $\omega$, with its stabilizer group $\mathds{Z}_3$, can give rise to mass matrix entries up to second-order terms, generating the desired charged lepton mass hierarchy using the smallness of $\varepsilon$.

We now provide a comprehensive analysis of the charged lepton mass hierarchy near the critical points $\ii$ and $\omega$, 
in the case where $L$ and $E^\ChargeC$ transform under a $\TpO$ reducible representation of $\Group{SL}{2,\mathds{Z}}$.
The corresponding weighted representations $\Omega$ and $\Omega_\ChargeC$ as defined in \Cref{eq:omegaDefinition} take the form of \Cref{eq:52a} and \Cref{eq:52b}.
After eliminating potential permutations, we are left with only eight inequivalent pairs of $(s_2,s_1)$ for the $\tau=\ii$ case, and seven inequivalent pairs of $(s_2,s_1)$ for the $\tau=\omega$ case. 
Since the singular values of the Dirac matrix do not change under the transpose operation, swapping the left- and right-handed fields, i.e.\ interchanging the representation assignments ($\Omega_\ChargeC \leftrightarrow \Omega$), results in an identical mass spectrum. Consequently, there are only 12 distinct unordered pairs of $(\Omega_\ChargeC,\Omega)$ that each can produce different rank 3 mass matrices for the $\tau=\ii$ case, which eventually lead to two different mass hierarchy structures as shown in \Cref{tab:chargedLeptonHierarchyI}. 
There are 28 distinct unordered pairs of $(\Omega_\ChargeC,\Omega)$ for the $\tau=\omega$ case, which ultimately yield six different mass hierarchy structures as summarized in \Cref{tab:chargedLeptonHierarchy}. 
Note that those cases where $\Omega=\Omega_\ChargeC$ can also be applied to the case of Majorana neutrinos originating from the Weinberg operator.

As we expected, near the point $\tau=\ii$, the hierarchical structure is at most $\varepsilon \relative 1 \relative 1$, which is consistent with neutrino masses with inverted ordering. Near the point $\omega$, we find that there are hierarchical structures $\varepsilon \relative 1 \relative 1$, $\varepsilon \relative \varepsilon \relative 1$, and $\varepsilon^2 \relative \varepsilon \relative 1$, which are respectively compatible with the neutrino masses with inverted ordering, the neutrino masses with normal ordering, and the charged lepton masses.

\section{Bounds on expansion coefficient for modular forms}
\label{app:Estimate}

In this appendix, we derive a rough upper bounds for the Taylor expansion coefficients of a holomorphic modular form $f(\tau) \in \mathcal{M}_k(\Gamma)$, where $ \Gamma \subseteq \Group{SL}{2,\mathds{Z}}$ is a normal subgroup with finite index, and $ k>0 $ is the weight. 
$f$ is holomorphic and defined on the upper half-plane $ \mathcal{H} = \{ \tau \in \mathds{C} \mid \im\tau > 0 \} $.
Consider a fixed point $ \tau_0 = x_0 + \ii y_0 \in \mathcal{H} $, and let $ \bar{\tau}_0 = x_0 - \ii y_0 $. Define the Cayley transformation
\begin{equation}
\varepsilon = \frac{\tau - \tau_0}{\tau - \bar{\tau}_0}
\end{equation}
mapping $\mathcal{H}$ to the unit disk $\mathcal{D}= \{ \varepsilon \in \mathds{C} \mid |\varepsilon| < 1 \} $. 
Then, $ f(\tau) $ admits a power series expansion, also known as the Taylor expansion,
\begin{equation}\label{eq:Taylor_expansion}
f\bigl(\varepsilon(\tau)\bigr) = \sum_{n=0}^{\infty} c_n\, \varepsilon^n\;.
\end{equation}
Using the Cauchy's estimate, the expansion coefficients $ c_n $ satisfy
\begin{equation}\label{eq:CauchyEstimate}
|c_n| \leq \frac{M_r}{r^n}\;,
\end{equation}
where $ 0 < r < 1 $, and $\displaystyle M_r =\max_{|\varepsilon| = r} \left\lvert f\bigl(\varepsilon(\tau)\bigr)\right\rvert$.

To proceed, we need to estimate the upper bound of $ M_r $. 
The contour $ |\varepsilon| = r $ corresponds to a curve $ \gamma_r $ in $ \mathcal{H} $, known as an Apollonius circle, and is defined by $ \left\lvert \frac{\tau - \tau_0}{\tau - \bar{\tau}_0} \right\rvert = r $. 
This circle lies within $ \mathcal{H} $ for $ r < 1 $. 
Now, we pull $ M_r$ from the circle $|\varepsilon|=r$ in $\mathcal{D}$ back to curve $\gamma_r$ in $\mathcal{H}$, 
\begin{equation}
  M_r =\max_{|\varepsilon| = r} \left\lvert f\bigl(\varepsilon(\tau)\bigr)\right\rvert = \sup_{\tau\in\gamma_r} \left\lvert f\bigl(\tau(\varepsilon)\bigr)\right\rvert
\end{equation} 
with $ \tau(\varepsilon) $ being the inverse transformation, 
\begin{equation}
 \tau(\varepsilon)= \frac{\bar{\tau}_0\, \varepsilon - \tau_0}{\varepsilon - 1}\;.
\end{equation} 
So the problem becomes the upper bound estimation of $|f(\tau)|$ on the curve $\gamma_r$. 
Let us recall the well known moderate growth conditions of modular forms in the upper half-plane $\mathcal{H}$~\cite{cohen2017modular},
\begin{equation}\label{eq:growth_of_f(tau)}
|f(\tau)| \leq 
 \begin{dcases}
  C\,(\im \tau)^{-k/2} &\text{for cusp forms ($ f \in \mathcal{S}_k(\Gamma) $)}\;,\\
  C \, \max\left\{1, ~(\im \tau)^{-k}\right\} & \text{for non cusp forms ($ f \in \mathcal{M}_k(\Gamma) \setminus \mathcal{S}_k(\Gamma) $)}\;.
 \end{dcases}
\end{equation}
Here $C$ is a constant. 
On the curve $ \gamma_r $, the imaginary part $ \im\tau $ has a lower bound $m_r \defeq y_0\, \frac{1 - r}{1 + r} \leq\im\tau $, where $ y_0 = \im\tau_0$. 
Applying the growth conditions \eqref{eq:growth_of_f(tau)}, we have 
\begin{equation}
M_r \leq \begin{dcases}
  C\, (m_r)^{-k/2} = C\, \left( y_0\, \frac{1 - r}{1 + r} \right)^{-k/2}
  & \text{for cusp forms}\;,\\
  C\, \max\left\{1, (m_r)^{-k}\right\} = C\, \max\left\{1, ~ \left( y_0\, \frac{1 - r}{1 + r} \right)^{-k}\right\}& \text{for non-cusp forms}\;.
 \end{dcases} 
\end{equation}
Using the Cauchy estimate \eqref{eq:CauchyEstimate} yields
\begin{equation}
 |c_n| \leq
 \begin{dcases}
  C\, y_0^{-k/2}\, \left( \frac{1 + r}{1 - r} \right)^{k/2}\, r^{-n}  
  &\text{for cusp forms}\;,\\ 
  C\, \max\left\{ r^{-n}, ~ y_0^{-k}\, \left( \frac{1 + r}{1 - r} \right)^k\, r^{-n}\right\} 
  &\text{for non-cusp forms}\;.
 \end{dcases}
\end{equation}

Choosing $ r = 1 - \frac{a}{n} $ optimizes the bound for large $ n $, where $a=k/2$ for cusp form and $a=k$ for non-cusp form,
\begin{equation}
 |c_n| \leq
 \begin{dcases}
  C \,\left(\frac{4\ee}{k\,y_0}\right)^{k/2}\, n^{k/2}
  &\text{for cusp forms}\;,\\
  C \,\left(\frac{2\ee}{k\,y_0}\right)^{k}\, n^k
  &\text{for non-cusp forms}\;.
 \end{dcases}  
\end{equation}
Thus, for large $n$, the growth of the Taylor coefficients $c_n$ in \Cref{eq:Taylor_expansion} are bounded according to the modular weight $k$, $ |c_n| =\mathcal{O}(n^{k/2}) $ for cusp forms and $|c_n| = \mathcal{O}(n^k) $ for non-cusp forms.

Note that the Fourier coefficients of $f(\tau)=\sum_{n=0}^{\infty} a_n \ee^{2\pi\ii n \tau}$ exhibit similarly simple bounds.
One has $|a_n| = \mathcal{O}(n^{k/2}) $ for cusp forms and $ |a_n| = \mathcal{O}(n^{k-1}) $ for non-cusp forms. 
This is also known as Hecke's bound~\cite{cohen2017modular}.

\bibliography{CLLR}
\bibliographystyle{utphys}

\begin{acronym}
 \acro{UV}{ultraviolet}
 \acro{CG}{Clebsch--Gordan} 
 \acro{EFT}{effective field theory}
 \acro{FI}{Fayet--Iliopoulos~\protect\cite{Fayet:1974jb}}
 \acro{FN}{Froggatt--Nielsen~\protect\cite{Froggatt:1978nt}}
 \acro{irrep}{irreducible representation}
 \acro{SM}{standard model of particle physics}
 \acro{SUGRA}{supergravity}
 \acro{SUSY}{supersymmetry}
 \acro{VEV}{vacuum expectation value}
 \acro{VVMF}{vector-valued modular form}
\end{acronym}

\end{document}